\documentclass[twocolumn]{aastex631}

\newcommand{\WHL}{WHL0137$-$08}
\newcommand{\MACS}{MACS0647+70}
\newcommand{\HST}{\textit{HST}}

\newcommand{\JWST}{\textit{JWST}}

\newcommand{\grizli}{\textsc{grizli}}
\newcommand{\eazypy}{\textsc{eazypy}}
\newcommand{\pixedfit}{\textsc{piXedfit}}
\newcommand{\photutils}{\textsc{photutils}}
\newcommand{\sextractor}{\textsc{SExtractor}}
\newcommand{\SEP}{\textsc{sep}}
\newcommand{\astropy}{\textsc{astropy}}
\newcommand{\astrodrizzle}{\textsc{astrodrizzle}}

\newcommand{\galfit}{\textsc{GALFIT}}

\newcommand{\mass}{$M_{*}$}
\newcommand{\massd}{$\Sigma_{*}$}
\newcommand{\sfrd}{$\Sigma_{\rm SFR}$}
\newcommand{\sigmakpc}{$\Sigma_{*,1\rm{kpc}}$}
\newcommand{\ssfrin}{$\text{sSFR}_{\rm in}$}
\newcommand{\ssfrout}{$\text{sSFR}_{\rm out}$}
\newcommand{\ratioradius}{$R_{e,\rm{SFR}}/R_{e}$}
\newcommand{\ratiossfr}{$\text{sSFR}_{\rm in}/\text{sSFR}_{\rm out}$}
\newcommand{\ssfrkpc}{$\text{sSFR}_{\rm 1kpc}$}

\submitjournal{ApJ}

\shorttitle{Galaxy Growth and Quenching Over Cosmic Time}
\shortauthors{Abdurro'uf et al.}

\begin{document}

\title{Spatially Resolved Stellar Populations of $0.3<z<6.0$ Galaxies in \WHL\ and \MACS\ Clusters as Revealed by JWST: How do Galaxies Grow and Quench Over Cosmic Time?}

\correspondingauthor{Abdurro'uf}
\email{fabdurr1@jhu.edu}

\author[0000-0002-5258-8761]{Abdurro'uf}
\affiliation{Center for Astrophysical Sciences, Department of Physics and Astronomy, The Johns Hopkins University, 3400 N Charles St. Baltimore, MD 21218, USA}
\affiliation{Space Telescope Science Institute (STScI), 
3700 San Martin Drive, Baltimore, MD 21218, USA}

\author[0000-0001-7410-7669]{Dan Coe}
\affiliation{Space Telescope Science Institute (STScI), 
3700 San Martin Drive, Baltimore, MD 21218, USA}
\affiliation{Association of Universities for Research in Astronomy (AURA), Inc.
for the European Space Agency (ESA)}
\affiliation{Center for Astrophysical Sciences, Department of Physics and Astronomy, The Johns Hopkins University, 3400 N Charles St. Baltimore, MD 21218, USA}

\author[0000-0003-1187-4240]{Intae Jung}
\affiliation{Space Telescope Science Institute (STScI), 3700 San Martin Drive, Baltimore, MD 21218, USA}

\author[0000-0001-7113-2738]{Henry C. Ferguson}
\affiliation{Space Telescope Science Institute (STScI), 3700 San Martin Drive, Baltimore, MD 21218, USA}

\author[0000-0003-2680-005X]{Gabriel Brammer}
\affiliation{Cosmic Dawn Center (DAWN), Copenhagen, Denmark}
\affiliation{Niels Bohr Institute, University of Copenhagen, Jagtvej 128, Copenhagen, Denmark}

\author[0000-0001-9298-3523]{Kartheik G. Iyer}
\altaffiliation{Hubble Fellow}
\affiliation{Columbia Astrophysics Laboratory, Columbia University, 550 West 120th Street, New York, NY 10027, USA}

\author[0000-0002-7908-9284]{Larry D. Bradley}
\affiliation{Space Telescope Science Institute (STScI), 3700 San Martin Drive, Baltimore, MD 21218, USA}

\author[0000-0001-8460-1564]{Pratika Dayal}
\affiliation{Kapteyn Astronomical Institute, University of Groningen, P.O. Box 800, 9700 AV Groningen, The Netherlands}

\author[0000-0001-8156-6281]{Rogier A. Windhorst} 
\affiliation{School of Earth and Space Exploration, Arizona State University, Tempe, AZ 85287-1404, USA}

\author[0000-0002-0350-4488]{Adi Zitrin}
\affiliation{Physics Department, Ben-Gurion University of the Negev, P.O. Box 653, Be'er-Sheva 84105, Israel}

\author[0000-0002-7876-4321]{Ashish Kumar Meena}
\affiliation{Physics Department, Ben-Gurion University of the Negev, P.O. Box 653, Be'er-Sheva 84105, Israel}

\author[0000-0003-3484-399X]{Masamune Oguri}
\affiliation{Center for Frontier Science, Chiba University, 1-33 Yayoi-cho, Inage-ku, Chiba 263-8522, Japan}
\affiliation{Department of Physics, Graduate School of Science, Chiba University, 1-33 Yayoi-Cho, Inage-Ku, Chiba 263-8522, Japan}

\author[0000-0001-9065-3926]{Jose M. Diego}
\affiliation{Instituto de F\'isica de Cantabria (CSIC-UC). Avda. Los Castros s/n. 39005 Santander, Spain}

\author[0000-0002-5588-9156]{Vasily Kokorev}
\affiliation{Kapteyn Astronomical Institute, University of Groningen, P.O. Box 800, 9700 AV Groningen, The Netherlands}

\author[0000-0001-7399-2854]{Paola Dimauro}
\affiliation{INAF - Osservatorio Astronomico di Roma, via di Frascati 33, 00078 Monte Porzio Catone, Italy}

\author[0000-0002-0786-7307]{Angela Adamo}
\affiliation{Department of Astronomy, Oskar Klein Centre, Stockholm University, AlbaNova University Centre, SE-106 91 Stockholm, Sweden}

\author[0000-0003-1949-7638]{Christopher J. Conselice}
\affiliation{Jodrell Bank Centre for Astrophysics, University of Manchester, Oxford Road, Manchester UK}

\author[0000-0003-1815-0114]{Brian Welch}
\affiliation{Department of Astronomy, University of Maryland, College Park, MD 20742, USA}
\affiliation{Observational Cosmology Lab, NASA Goddard Space Flight Center, Greenbelt, MD 20771, USA}
\affiliation{Center for Research and Exploration in Space Science and Technology, NASA/GSFC, Greenbelt, MD 20771}

\author[0000-0002-5057-135X]{Eros Vanzella}
\affiliation{INAF -- OAS, Osservatorio di Astrofisica e Scienza dello Spazio di Bologna, via Gobetti 93/3, I-40129 Bologna, Italy}

\author[0000-0003-4512-8705]{Tiger Yu-Yang Hsiao}
\affiliation{Center for Astrophysical Sciences, Department of Physics and Astronomy, The Johns Hopkins University, 3400 N Charles St. Baltimore, MD 21218, USA}

\author[0000-0002-9217-7051]{Xinfeng Xu}
\affiliation{Center for Astrophysical Sciences, Department of Physics and Astronomy, The Johns Hopkins University, 3400 N Charles St. Baltimore, MD 21218, USA}

\author[0000-0002-4430-8846]{Namrata Roy}
\affiliation{Center for Astrophysical Sciences, Department of Physics and Astronomy, The Johns Hopkins University, 3400 N Charles St. Baltimore, MD 21218, USA}

\author[0000-0003-2434-2657]{Celia R. Mulcahey}
\affiliation{Center for Astrophysical Sciences, Department of Physics and Astronomy, The Johns Hopkins University, 3400 N Charles St. Baltimore, MD 21218, USA}




\begin{abstract}

We study the spatially resolved stellar populations of 444 galaxies at $0.3<z<6.0$ in two clusters (WHL0137-08 and MACS0647+70) and a blank field, combining imaging data from \HST\ and \JWST\ to perform spatially resolved spectral energy distribution (SED) modeling using \pixedfit. The high spatial resolution of the imaging data combined with magnification from gravitational lensing in the cluster fields allows us to resolve a large fraction of our galaxies (109) to sub-kpc scales. At redshifts around cosmic noon and higher ($2.5\lesssim z\lesssim 6.0$), we find mass doubling times to be independent of radius, inferred from flat specific star formation rate (sSFR) radial profiles and similarities between the half-mass and half-SFR radii. At lower redshifts ($1.5\lesssim z\lesssim 2.5$), a significant fraction of our star-forming galaxies show evidence for nuclear starbursts, inferred from centrally elevated sSFR, and a much smaller half-SFR radius compared to the half-mass radius. At later epochs, we find more galaxies suppress star formation in their centers but are still actively forming stars in the disk. Overall, these trends point toward a picture of inside-out galaxy growth consistent with theoretical models and simulations. We also observe a tight relationship between the central mass surface density and global stellar mass with $\sim 0.38$ dex scatter. Our analysis demonstrates the potential of spatially resolved SED analysis with \JWST\ data. Future analysis with larger samples will be able to further explore the assembly of galaxy mass and the growth of their structures. 
\end{abstract}



\section{Introduction} \label{sec:intro}

Over the last few decades, multiwavelength studies of galaxies throughout cosmic history reveal that the global star formation rate density (SFRD) in the universe was increasing with cosmic time from the reionization epoch and reached a peak at $z\sim 2$ ($\sim 3.5$ Gyr after the Big Bang; cosmic noon) after which it declined exponentially toward the present day \citep{2014Madau}. In this picture, it is estimated that $\sim$25\% of the present-day stellar mass density (SMD) was formed before the peak of the cosmic SFRD, around half of the SMD was formed during $0.7<z<2.0$, and another $\sim$25\% was formed since $z=0.7$ \citep[i.e., around the last half of the universe's age;][]{2014Madau}. Although the cosmic SFRD at early cosmic time is still debated due to the dust obscuration effects (see e.g.,~\citealt{2021Fudamoto, 2018Casey}), an emerging picture is that cosmic SMD increases with cosmic time since the epoch of reionization, which is believed to take place before $z\sim 6$ \citep[e.g.,][]{2013Treu,2015McGreer,2018Dayal}. 

Observations also revealed that most of the star formation occurs in galaxies that lie in the so-called star-forming main sequence (SFMS), which is a tight nearly linear correlation between the integrated (i.e.,~global) star formation rate (SFR) and stellar mass \citep[$M_{*}$;][]{2004Brinchmann, 2007Daddi, 2007Elbaz, 2007Noeske, 2012Whitaker, 2014Whitaker, 2014Speagle, 2015Salmon, 2016Tomczak, 2017Santini, 2018Iyer, 2020Leslie, 2022Leja, 2023Popesso}. This relation has been shown to hold at any epoch with a nearly constant scatter \citep[$\sim 0.3$ dex;][]{2012Whitaker,2014Speagle,2023Popesso}, suggesting that galaxies grow in mass over cosmic time in a state of self-regulated semi-equilibrium \cite[e.g.,][]{2010Bouche, 2010Daddi, 2010Genzel, 2016Tacchella, 2020Tacchella}. Understanding this process in detail and the mechanisms that shut down star formation in galaxies and move them out of the SFMS onto the ``quenched'' population requires knowledge of not only integrated galaxy properties but also spatially resolved structures within galaxies. 

The study of spatially resolved properties of galaxies with integral field spectroscopy (IFS) and high spatial resolution imaging has been providing important insights toward a better understanding of galaxy evolution. Among the important findings is the realization that some of the well-known scaling relations observed on global scales originated from similar relations on kpc scales within galaxies \citep[see a review by][]{2020Sanchez}. This includes the spatially resolved star-forming main sequence (rSFMS), a relationship between the SFR surface density ($\Sigma_{\rm SFR}$) and \mass\ surface density ($\Sigma_{*}$), a local analog of the global SFMS \citep[e.g.,][]{2013Sanchez, 2013Wuyts, 2016Cano-Diaz, 2017Abdurrouf, 2017Hsieh, 2018Abdurrouf, 2019Lin, 2020Enia}. Recent studies found that the rSFMS relation (and hence the global SFMS) originated from two more fundamental relations on kpc scales: the resolved Kennicutt--Schmidt ($\Sigma_{\rm SFR}$ versus $\rm{H}_{2}$ mass surface density; $\Sigma_{\rm{H}_{2}}$) and molecular gas main-sequence relations ($\Sigma_{*}$ versus $\Sigma_{\rm{H}_{2}}$) \citep[e.g.,][]{2019Lin, 2022Baker, 2022Abdurrouf2, 2020Morselli}. This emphasizes the necessity of studying the spatially resolved properties of galaxies.

Spatially resolved studies of high redshift galaxies ($z \sim 1-4$) have hinted at how galaxies assembled their structures. The emerging picture from these studies is that galaxies grow their mass in an inside-to-outside manner \citep[i.e., \textit{inside-out growth} scenario; e.g.,][]{2013vanDokkum, 2012Nelson, 2015Morishita, 2016Nelson} and cease their star formation activities in a similar manner \citep[i.e., \textit{inside-out quenching} scenario; e.g.,][]{2015Tacchella, 2017Jung, 2018Abdurrouf, 2018Tacchella, 2018Ellison, 2020Bluck}. \citet{2016Nelson} analyzed the spatially resolved distributions (on kpc scales) of $\text{H}\alpha$ emission and stellar mass of $0.7<z<1.5$ galaxies using the Hubble Space Telescope (\HST)/WFC3 grism data from the 3D-HST survey \citep{2014Skelton}. They found that the spatial distribution of $H_{\alpha}$ emission in the galaxies is more extended than the stellar mass distribution, suggesting that the past star formation in the galaxies has accumulated stellar mass in the center and now the star formation progresses outward to assemble the disk. \citet{2015Tacchella} analyzed the spatial distributions of SFR and $M_{*}$ of $\sim 30$ star-forming galaxies at $z\sim 2$ using IFS data from the SINS/zC-SINF survey \citep{2018ForsterSchreiber}. They observed that massive galaxies ($M_{*}\gtrsim 10^{11}M_{\odot}$) in their sample have a centrally-suppressed specific SFR (sSFR) radial profile and a massive central spheroid that is as dense as the centers of local early-type galaxies. In contrast to this, less massive galaxies in their sample have broadly flat sSFR radial profiles. This trend indicates that massive galaxies at this epoch might have started a quenching process in their central regions and assembled a mature bulge.      

The buildup of the central stellar mass density is likely correlated with the quenching process in galaxies. The central stellar mass density within a 1 kpc radius (\sigmakpc) has been shown to be a good predictor for quiescence, where galaxies with high \sigmakpc\ tend to be red and quiescent, whereas galaxies with low \sigmakpc\ tend to be blue and star-forming \citep[e.g.,][]{2013Fang,2015Tacchella,2016Tacchella2, 2017Barro, 2017Jung, 2017Whitaker, 2022Dimauro}. It has also been shown that \sigmakpc\ is tightly correlated with the global \mass, suggesting that \mass\ of galaxies grow hand-in-hand with the central mass density. In this \sigmakpc--\mass\ relation, quiescent galaxies reside in a sequence at the tip of the overall relationship and have a shallower slope than the relation with star-forming galaxies only, indicating a formation of a matured bulge in the quiescent galaxies \citep{2013Fang,2015Tacchella,2017Barro}. 

Galaxies also grow their sizes hand-in-hand with the global \mass, as indicated by the size--mass relation \citep[e.g.,][]{2003Shen,2014vanderWel,2019Suess}. Previous studies have shown that star-forming and quiescent galaxies follow very different size--mass relations where quiescent galaxies tend to be more compact (i.e.,~having smaller size) in all \mass\ and exhibit steeper relation than the star-forming galaxies \citep{2014vanderWel,2021Yang}. A possible explanation for this trend is that star-forming galaxies build their mass at all radii by mostly in-situ star formation, whereas quiescent galaxies grow inside-out through mergers \citep[e.g.,][]{2015vanDokkum}.     

Previous studies, some of which are mentioned above, have used \HST\ for resolving galaxies out to $z\sim 3$, roughly a limit where galaxies can be resolved well by the telescope, given its spatial resolution and depth. Furthermore, the wavelength coverage of \HST\ only covers the rest-frame ultraviolet (UV) and a small portion of the optical at $z\sim 3$, making it difficult to robustly derive \mass\ as well as the other stellar population properties, which typically requires a rest-frame near-infrared (NIR). Forcing to include NIR imaging from the ground-based telescopes would need to sacrifice the spatial resolution of \HST\ \citep[e.g.,][]{2017Jung}. With the advent of the James Webb Space Telescope (\JWST) NIRCam observations \citep{2022Rieke, 2022Rigby}, with its high spatial resolution, depth, and its coverage in NIR, now we can push the analysis of spatially resolved SED of galaxies to higher redshifts. Some very recent studies have used \JWST/NIRCam imaging to study the internal structures and morphology of galaxies at $z>3$ \citep[e.g.,][]{2022Ferreira, 2022Chen, 2022Kartaltepe, 2022PerezGonzalez, 2022GimenezArteaga}, and even resolving a lensed galaxy at $z\sim 11$ \citep{2022Hsiao}. 

In this paper, we use imaging data from \HST/ACS and \JWST/NIRCam to analyze the spatially resolved SEDs of $0.3<z<6.0$ galaxies in the sightlines of WHLJ013719.8--082841 (hereafter \WHL; RA = 01:37:25.0, DEC = $-$08:27:23, J2000; $z = 0.566$; \citealt{2012Wen, 2015Wen}) and MACSJ0647.7+7015 (hereafter \MACS; RA = 06:47:50.03, DEC = $+$70:14:49.7, J2000; $z=0.591$; \citealt{2007Ebeling}) clusters and examine the spatial distributions of their stellar populations. Our main goal is to get hints on the assembly of galaxy structures over cosmic time, especially how galaxies build their stellar masses and quench their star formation activities. The high spatial resolution of \JWST/NIRCam combined with magnification from gravitational lensing in the cluster fields, allows us to resolve high-redshift galaxies down to sub-kpc scales. Our method using \pixedfit\ \citep{2022Abdurrouf3} can simultaneously process imaging data, perform pixel binning to optimize the signal-to-noise (S/N) ratio of the spatially resolved SEDs, and perform SED fitting. The wavelength coverage of \HST/ACS and \JWST/NIRCam allow us to get full coverage of the rest-frame UV to NIR for the majority of our sample, which can give a strong constraint on model SEDs and break the age--dust--metallicity degeneracy (see Appendix~\ref{sec:breaking_degeneracies}). While IFS observation at $z\gtrsim 2$ is lacking, our analysis in this paper provides a good alternative for the analysis of spatially resolved SED of high redshift galaxies. Our analysis in this paper is one of the first robust spatially resolved SED analyses of hundreds of galaxies using \JWST\ data. \citet{2021Abdurrouf} have demonstrated the capabilities of spatially resolved SED fitting using \pixedfit\ on local galaxies. In particular, it gives robust SFR on kpc scales when rest-frame UV--NIR photometry is available, which is consistent with the SFR derived from $\text{H}\alpha$ emission maps (dust-corrected based on the Balmer decrement) from the MaNGA IFS survey \citep{2015Bundy}.  

The paper is organized as follows. In Section~\ref{sec:data_sample}, we present the data and sample galaxies. We describe the spatially resolved SED fitting methodology in Section~\ref{sec:method} and present our results in Section~\ref{sec:results}, which include the radial profiles of some key stellar population properties, comparison between the compactness of the spatial distributions of SFR and \mass, and \sigmakpc--\mass\ relation. In Section~\ref{sec:discussions}, we further discuss our results, focusing on the evolutionary trends with redshift and the implications to the study of galaxy evolution. 

Throughout this paper, we assume the \citet{2003Chabrier} initial mass function (IMF) with a mass range of $0.1-100M_{\odot}$ and cosmological parameters of $\Omega_{m}=0.3$, $\Omega_{\Lambda}=0.7$, and $H_{0}=70\text{ km}\text{ s}^{-1}\text{ Mpc}^{-1}$.

\section{Data and Sample} \label{sec:data_sample}

\subsection{Observational Data} \label{sec:data}

\begin{deluxetable*}{cccccccc}
\tablecaption{\HST\ and \JWST\ Imaging Data Used in the Spatially Resolved SED Fitting \label{tab:obs_data}}
\tablewidth{\columnwidth}
\tablehead{
\colhead{Telescope} &
\colhead{Camera} &
\colhead{Filter} &
\colhead{Wavelength} &
\multicolumn{2}{c}{Depth\textsuperscript{a}} &
\multicolumn{2}{c}{PSF FWHM\textsuperscript{b}}
\\[-3pt]
\cline{5-8}
\colhead{} &
\colhead{} &
\colhead{} &
\colhead{} &
\colhead{\WHL} &
\colhead{\MACS} &
\colhead{\WHL} &
\colhead{\MACS}
\\[-6pt]
\colhead{} &
\colhead{} &
\colhead{} &
\colhead{(\micron)} &
\colhead{(AB mag)} &
\colhead{(AB mag)} &
\colhead{(arcsec)} &
\colhead{(arcsec)}
}
\startdata
\HST\ & ACS/WFC & F435W & 0.37--0.47 & 27.7 & 28.0 & 0.11 & 0.11 \\
\HST\ & ACS/WFC & F475W & 0.4--0.55 & 28.5 & 28.2 & 0.11 & 0.11 \\
\HST\ & ACS/WFC & F555W & 0.46--0.62 & \nodata & 28.7 & \nodata & 0.11 \\
\HST\ & ACS/WFC & F606W & 0.47--0.7 & 28.3 & 28.3 & 0.11 & 0.11 \\
\HST\ & ACS/WFC & F625W & 0.54--0.71 & \nodata & 27.9 & \nodata & 0.11 \\
\HST\ & ACS/WFC & F775W & 0.68--0.86 & \nodata & 27.8 & \nodata & 0.08 \\
\HST\ & ACS/WFC & F814W & 0.7--0.95 & 28.7 & 28.5 & 0.11 & 0.11 \\
\JWST\ & NIRCam & F090W & 0.8--1.0 & 28.3 & \nodata & 0.04 & \nodata \\
\JWST\ & NIRCam & F115W & 1.0--1.3 & 28.4 & 28.1 & 0.04 & 0.04 \\
\JWST\ & NIRCam & F150W & 1.3--1.7 & 28.5 & 28.3 & 0.06 & 0.06 \\
\JWST\ & NIRCam & F200W & 1.7--2.2 & 28.7 & 28.4 & 0.06 & 0.06 \\
\JWST\ & NIRCam & F277W & 2.4--3.1 & 29.1 & 28.9 & 0.11 & 0.11 \\
\JWST\ & NIRCam & F356W & 3.1--4.0 & 29.3 & 29.0 & 0.11 & 0.11 \\
\JWST\ & NIRCam & F410M & 3.8--4.3 & 28.6 & \nodata & 0.16 & \nodata \\
\JWST\ & NIRCam & F444W & 3.8--5.0 & 29.0 & 28.8 & 0.16 & 0.16
\enddata
\tablecomments{
\textsuperscript{a}$5\sigma$ point source AB magnitude limit measured within a $0\farcs2$ diameter circular aperture. 
\textsuperscript{b}PSF FWHM is based on empirical measurement as described in Appendix~\ref{sec:generate_psfs}.
}
\end{deluxetable*}
 
\vspace*{-2em}

\subsubsection{JWST Observations} \label{sec:data_jwst}

We obtain \JWST/NIRCam imaging data of \WHL\ cluster from Cycle 1 General Observers (GO) 2282 program (PI Coe) and \MACS\ cluster from GO 1433 program (PI Coe). The \WHL\ cluster was observed in July 2022, while the \MACS\ cluster was observed on 23 September 2022. The GO 2282 program aims at further investigating Earendel \citep{2022Welch1,2022Welch2} and the Sunrise Arc \citep{2022Vanzella}. The \JWST/NIRCam data from this program consist of eight filters (F090W, F115W, F150W, F200W, F277W, F365W, F410M, and F444W) spanning a wavelength range of $0.8-5.0~\micron$. The GO 1433 program is intended to observe the triply-lensed galaxy MACS0647–JD at $z\sim 11$ \citep{2013Coe, 2022Hsiao}. This program obtained \JWST/NIRCam imaging in six filters (F115W, F150W, F200W, F277W, F365W, and F444W) spanning $1-5~\micron$. The exposure time of each filter in the two programs is $2104$ seconds. It achieves $5\sigma$ limiting AB magnitude of $28.0$ to $29.0$ in a $r=0\farcs2$ diameter circular aperture. 

For each filter, we obtained four dithers using INTRAMODULEBOX primary dithers to cover the $4-5\arcsec$ gap between the sort wavelength (SW; $\lambda<2.4\micron$) detectors, improve the spatial resolution of final drizzled images, and minimize the impact of image artifacts and bad pixels. In each observation, we obtained NIRCam imaging over two $2\farcm26 \times 2\farcm26$ fields separated by 40\farcm5, covering a total area of 10.2 arcmin$^{2}$. In the observation of \WHL\ cluster, the NIRCam module B was centered at the cluster while module A covered a nearby field centered $\sim$ 2\farcm9 from the cluster center (hereafter called ``blank field''). On the other hand, the \MACS\ cluster was centered at module A, and module B observes a blank field nearby to it.

\subsubsection{HST Data} \label{sec:data_hst}

We obtain \HST\ imaging data of the \WHL\ cluster from the Reionization Lensing Cluster Survey (RELICS) \HST\ Treasury program \citep[GO 14096;][]{2019Coe}. The RELICS program obtained the first \HST\ imaging of the \WHL\ cluster in 2016 with three orbits of ACS (F435W, F606W, and F814W) and two orbits of WFC3/IR (F105W, F125W, F140W, and F160W) data spanning $0.4 - 1.7~\micron$. Two follow-up \HST\ imaging programs (GO 15842 and GO 16668; PI: Coe) have thus far obtained an additional 5 orbits of \HST\ ACS imaging in F814W, 2 orbits in F475W, and 4 orbits with WFC3/IR in F110W.

The \HST\ imaging data of the \MACS\ cluster are taken from multiple programs. Overall, \MACS\ has been observed in total of 39 orbits of \HST\ imaging in 17 filters. The cluster was first observed by programs GO 9722 (PI Ebeling) and GO 10493, 10793 (PI Gal-Yam) in the ACS F555W and F814W filters. Then additional imaging in 15 filters (WFC3/UVIS, ACS, and WFC3/IR, spanning $0.2-1.7~\micron$ ) was obtained by the Cluster Lensing and Supernova Survey with Hubble (CLASH; \citealt{2012Postman}; GO 12101, PI Postman). Finally, additional imaging in WFC3/IR F140W was obtained as part of a grism spectroscopy program (GO 13317, PI Coe). 

It is important to note that the nearby blank fields to the \WHL\ and \MACS\ clusters that were observed with NIRCam are not covered in the \HST\ observations described above. In this work, we analyze galaxies in three fields: \WHL\ cluster field, \MACS\ cluster, and the NIRCam blank field nearby to the \WHL\ (hereafter simply called blank field). We do not analyze galaxies in the NIRCam blank field of \MACS\ because it is observed in fewer filters than the blank field of \WHL\ and it does not have F090W observation, which prevents us from selecting galaxies at $z<2$ in this field as their photometry do not cover the rest-frame $4000$~\AA\ break. For the \WHL\ and the blank field, we use 4 \HST/ACS filters (F435W, F475W, F606W, and F814W) and 8 \JWST/NIRCam filters, whereas for the \MACS, we use 7 \HST/ACS filters (F435W, F475W, F555W, F606W, F625W, F775W, and F814W) and 6 \JWST/NIRCam filters. We do not use \HST/WFC3 IR filters to get high spatial resolution possible while still getting sufficiently wide wavelength coverage with the \HST/ACS and \JWST/NIRCam. Please refer to Table~\ref{tab:obs_data} for information on limiting magnitudes and the point spread function (PSF) sizes of our data.    

\subsection{Sample Galaxies} \label{sec:sample_selection}

We use \grizli\ v4 photometric catalogs (will be described in Section~\ref{sec:data_reduction}) to select our sample galaxies in the three fields (\WHL, blank field, and \MACS). The catalogs provide aperture fluxes and photometric redshifts with which we select our sample. The sample selection is described in the following. First, we select galaxies that have integrated signal-to-noise (S/N) ratio $>5$ in all \JWST\ filters that are available for the fields. This is to ensure that we will have galaxies with good photometry in at least \JWST\ filters. This initial cut selects 1322 (out of 2718), 1278 (out of 3032), and 1331 (out of 2660) galaxies in the \WHL, blank field, and \MACS, respectively. We do not apply the same S/N criteria on \HST\ filters because it would exclude more galaxies as they have lower S/N than \JWST\ filters.     

We further cut the sample galaxies based on their redshift to get a sufficient coverage of the rest-frame UV--NIR. For galaxies in the \WHL\ and \MACS, which are observed by both \JWST\ and \HST, we select galaxies at $0.3<z<6.0$, whereas for galaxies in the blank field, which do not have \HST\ observations, we select galaxies at $1.3<z<6.0$. This redshift cut ensures that the rest-frame $4000$~\AA\ break is covered. This cut further reduces the sample to 1258, 581, and 1257 galaxies in the \WHL, blank field, and \MACS, respectively. After that, we do a visual inspection to exclude galaxies that appear to be very small (i.e.,~unresolved) and in a merger (i.e.,~one segmentation region having multiple cores or multiple galaxies in one segmentation region, despite possible interlopers). This further reduces the sample to 354, 239, and 220 galaxies in the \WHL, blank field, and \MACS, respectively. 

\begin{figure}
\centering
\includegraphics[width=0.5\textwidth]{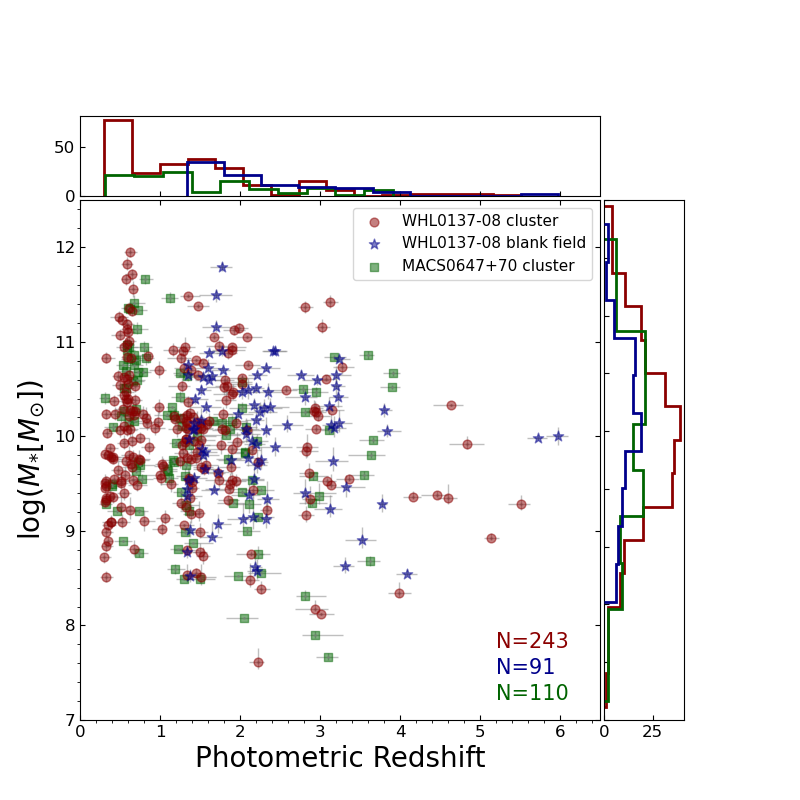}
\caption{Distributions of $M_{*}$ and redshifts of the sample galaxies analyzed in this work, which consists of galaxies in two cluster fields (\WHL\ and \MACS) and a blank field (a nearby field centered $\sim$ 2\farcm9 from the \WHL\ that was observed with \JWST/NIRCam). The global $M_{*}$ shown here are derived by summing up $M_{*}$ of pixels (in the galaxy's region) obtained from our spatially resolved SED fitting.}
\label{fig:redshift_vs_sm}
\end{figure}

We perform spatially resolved SED analysis on the galaxies in this initial sample. A detailed description of the methodology will be given in Section~\ref{sec:method}. Once the analysis is done, we inspect the results of all the galaxies and further exclude galaxies that seem to have bad SED fitting results based on the $\chi^{2}$ values of the fitting to the integrated SEDs within the central effective radius (see Section~\ref{sec:resolved_sedfit}) and the average $\chi^{2}$ values of the fitting to the first 20 spatial bins (see Section~\ref{sec:pixbin} for the definition of the spatial bin). We exclude galaxies that have $\chi^{2}>20$ for the SEDs within the central effective radius and average $\chi^{2}>40$ for the first 20 spatial bins. We note that $\chi^{2}$ value can be unrealistically high if systematic uncertainties of the photometry are not properly accounted for. Besides this, there is still uncertainty around the zero-point calibration of NIRCam photometry in the current early observations \citep[e.g.,][]{2022Boyer, 2022Finkelstein}. Therefore, we visually inspect the SED fitting results of each galaxy using similar plots as shown in Figure~\ref{fig:plot_sedfits}. We find that in most cases, NIRCam fluxes are well fitted by our models, better than \HST/ACS fluxes. This might be due to the shallower depths (and lower S/N) of \HST\ compared to \JWST. The $\chi^{2}$ values above are high enough to get a sufficient number of galaxies and low enough to get good quality SED fitting results. This results in our final sample, consisting of 243, 91, and 110 galaxies in the \WHL, blank field, and \MACS, respectively. Figure~\ref{fig:redshift_vs_sm} shows the distributions of redshifts and $M_{*}$ of our sample galaxies. 

We note that our sample selection may possibly bias toward selecting relatively massive, bright, and resolved (i.e.,~big) galaxies in each redshift. However, due to the lensing magnification in the cluster fields, we expect to detect on average lower-mass galaxies with better spatial resolution than in the normal fields. The small number of galaxies and the limited volume sampled might make our sample to be not representative of the general population of galaxies. However, since we do not make inferences on the average trends or number densities as a function of global properties (e.g.,~\mass), but instead we show trends in individual galaxies, our results still provide useful insights on the evolution of galaxy structures. We also ignore the possible contamination by the Active Galactic Nucleus (AGN) host galaxies in our current study because of the lack of diagnostics for identifying them using our current data.             

\section{Methodology} \label{sec:method}

\subsection{Data Reduction and Photometric Catalog} \label{sec:data_reduction}

We use the \grizli\ pipeline \citep{Grizli} to process the \HST\ FLT and the \JWST\ pipeline-calibrated level-2 imaging data. The \JWST\ data were processed using the calibration pipeline v1.5.3 with CRDS context \texttt{jwst\_0942.pmap}, which includes photometric calibrations based on in-flight data. The \JWST\ level-2 imaging data were then scaled with detector-dependent factors \citep{gabriel_brammer_2022_7143382} based on a NIRCam flux calibration using the standard star J1743045. Our photometric zeropoints described here are similar to those obtained by the \JWST\ Resolved Stellar Populations ERS program \citep{2022Boyer, 2022Nardiello} who analyzed the M92 globular cluster. We also check the consistency of our calibration with the more recent one based on CAL program data \texttt{jwst\_0989.pmap} and find out that they are consistent within 3\% in all filters analyzed here.

In processing the \JWST\ data, the \grizli\ pipeline applies a correction to reduce the effect of $1/f$ noise and masks ``snowballs''\footnote{\url{https://jwst-docs.stsci.edu/data-artifacts-and-features/snowballs-artifact}} effect caused by the large cosmic ray impacts to the NIRCam detectors. Besides this, the \grizli\ pipeline also corrects for the ``wisps''\footnote{\url{https://jwst-docs.stsci.edu/jwst-near-infrared-camera/nircam-features-and-caveats/nircam-claws-and-wisps}}, which is a faint, diffuse stray light features that appear at the same detector locations in NIRCam images and most prominent in the A3, B3, and B4 detectors in the F150W and F200W images. 

The \grizli\ pipeline aligns the \HST\ and \JWST\ imaging data to a common world coordinate system which is registered based on the GAIA DR3 catalogs \citep{Gaia_EDR3}. The images are then drizzled to a common pixel grid using the \astrodrizzle\ \citep{MultiDrizzle,
DrizzlePac}. The 17 \HST\ filters and 4 \JWST\ NIRCam long-wavelength (LW) filters (F277W, F356W, F410M, and F444W) are drizzled to a spatial sampling of 0$\farcs$04 per pixel while the \JWST\ short-wavelength (SW) filters (F090W, F115W, F150W, and F200W) are drizzled to a spatial sampling of 0$\farcs$02 per pixel. 

Source detection is then performed on a weighted sum of the drizzled NIRCam images in all filters using \SEP\ \citep{SEP, 1996Bertin}. Fluxes are then calculated for each source in three circular apertures, $0\farcs36$, $0\farcs5$, and $0\farcs7$. Then photometric redshift measurement is performed using the $0\farcs5$ aperture SEDs employing \eazypy \citep{2008Brammer}. This code fits observed photometry using a set of templates added in a non-negative linear combination. The processed imaging data along with the photometric catalog are publicly available\footnote{\url{https://cosmic-spring.github.io/index.html}}. These data products have also been used in some recent studies \citep{2022Welch2, 2022Bradley, 2022Hsiao, 2022Vanzella, 2022Meena}. 

\subsection{Analysis of Post-processed Imaging Data} \label{sec:images_analysis}

In this work, we combine the post-processed \HST\ and \JWST\ imaging data (in up to 13 filters) into a common spatial resolution (i.e.,~PSF size) and sampling (i.e.,~pixel size) for extracting the spatially resolved SEDs of our sample galaxies. These spatially resolved SEDs are then fitted with models to infer the underlying properties of the stellar populations. We use \pixedfit\footnote{\url{https://github.com/aabdurrouf/piXedfit}} \citep{2021Abdurrouf, 2022Abdurrouf3} throughout this analysis. Basically, this process includes three main tasks: image processing, pixel binning, and SED fitting. We will briefly describe these steps in the following. 

The image processing is carried out automatically using \pixedfit. For each galaxy, we first crop stamp images with a size of 6$\farcs$04 $\times$ 6$\farcs$04 (corresponding to 302$\times$302 pixels in NIRCam SW and 151$\times$151 pixels in NIRCam LW and \HST/ACS filters) centered at the galaxy. We then perform background subtraction to each stamp image using \photutils\ \citep{Bradley2022b}. Next, we perform point spread function (PSF) matching to homogenize the spatial resolution across filters. We degrade the spatial resolution of the images to match the resolution of the F444W filter, which has the lowest spatial resolution (see Table~\ref{tab:obs_data}). For this, we generate the empirical PSFs of \HST/ACS and \JWST/NIRcam filters along with the convolution kernels using \photutils\ package (see Appendix~\ref{sec:generate_psfs}). The PSF matching is carried out by convolving the stamp images with the convolution kernels. After PSF matching, we register all the stamp images to a common spatial sampling of 0$\farcs$04 per pixel. At the end, we have multiband stamp images with a size of 151$\times$151 pixels for each galaxy in our sample.

\subsection{Constructing Photometric Data Cubes} \label{sec:generate_photo_data_cubes}

\pixedfit\ further processes the stamp images to produce photometric data cubes. First, it defines a galaxy's region of interest. For each galaxy, segmentation maps are first produced in all filters using \SEP\ \citep{SEP} and those maps are then merged together into a single map. In the segmentation process, we use the same parameters for all filters as follows. We set the detection threshold (\texttt{thresh}), the number of thresholds for deblending (\texttt{deblend\_nthresh}), and the minimum contrast ratio for deblending (\texttt{deblend\_cont}) to be $2.0$, $40$, and $0.005$, respectively.

In some cases, the merged segmentation map is larger than expected, as can be inferred from the maps of multiband fluxes. This can be caused by some factors, for example, interference from neighboring objects that are not separated well by the deblending process. We visually inspect the merged segmentation map of each galaxy to find out this issue. To deal with this, we tweak the deblending parameters to get cleaner segmentation maps or ignore the segmentation map in some filters that have this deblending issue, then merge them again. 

\begin{figure*}
\centering
\includegraphics[width=0.9\textwidth]{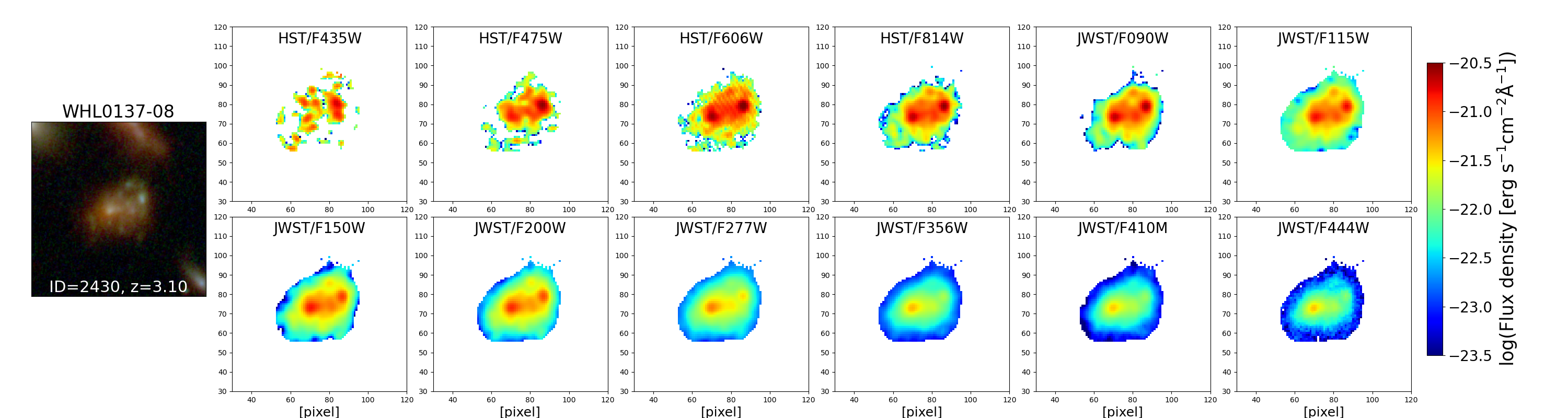}
\includegraphics[width=0.7\textwidth]{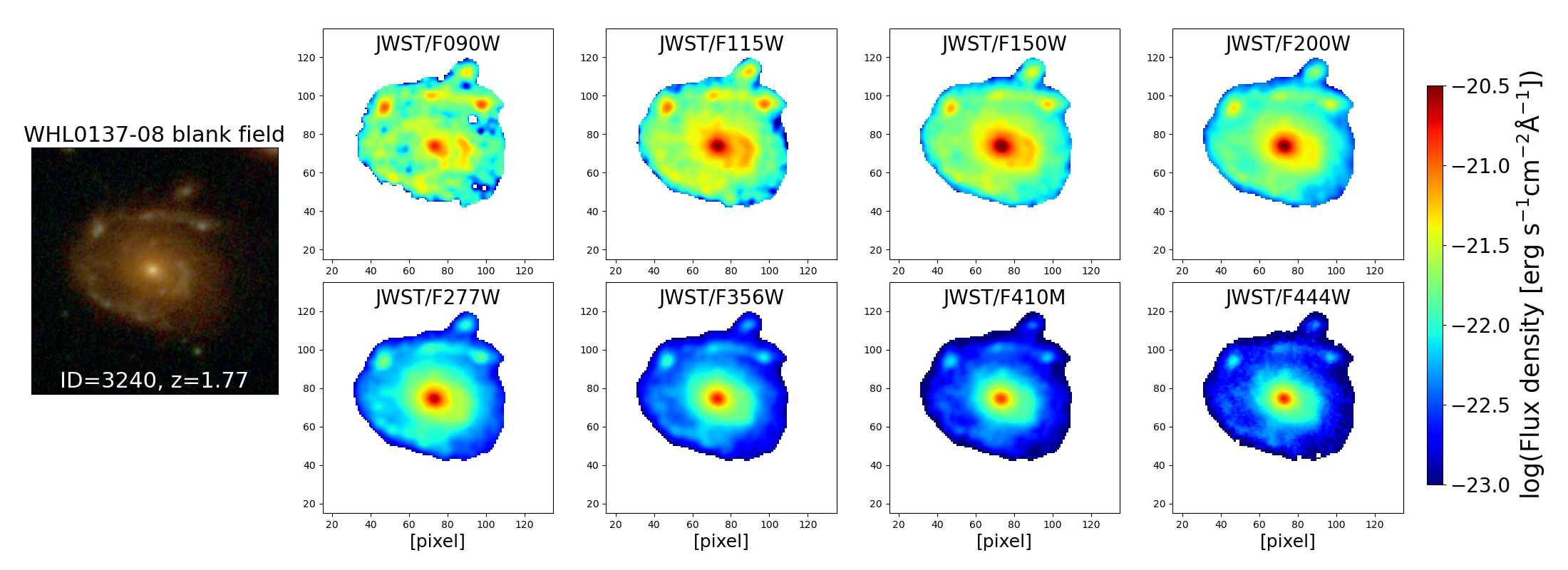}
\includegraphics[width=0.99\textwidth]{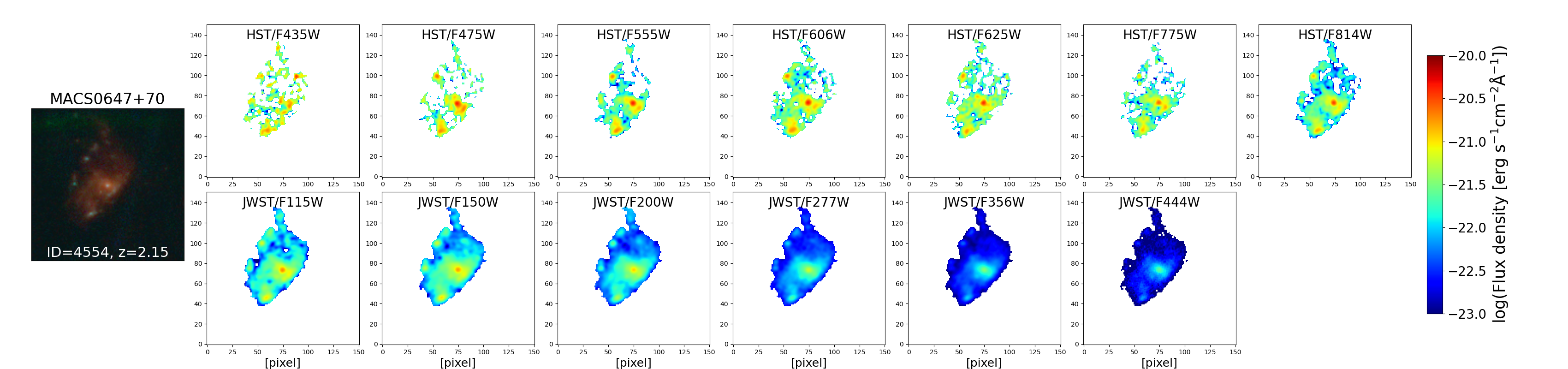}
\caption{Examples of the maps of multiband fluxes produced from the image processing. An example of one galaxy is shown for each field, from top to bottom: \WHL\ (observed in 12 filters), blank field (8 filters), and \MACS\ (13 filters). The galaxy ID is based on our \grizli\ v4 public catalog.}
\label{fig:img_process}
\end{figure*}

Once the galaxy's region is defined, then the fluxes of pixels within the region are calculated. We use the PHOTFLAM keyword in the header of the \grizli\ imaging data products to convert the pixel value into flux density in the units of erg $\text{s}^{-1}\text{cm}^{-2}$\AA$^{-1}$. The data cubes are then stored in FITS files. Figure~\ref{fig:img_process} shows examples of the maps of multiband fluxes of galaxies in three fields analyzed in this work. The color images shown in the leftmost panels are created using \texttt{Trilogy}\footnote{\url{https://github.com/dancoe/trilogy}} \citep{2012Coe}. 

\subsection{Pixel binning} \label{sec:pixbin}

The SEDs of pixels are usually noisy and might not provide sufficient constraint to the models if the fitting is performed to them. Therefore, we perform pixel binning using \pixedfit\ to optimize the signal-to-noise ratio (S/N) of the spatially resolved SEDs. Basically, this process bins neighboring pixels to achieve a certain S/N ratio threshold that can be set in multiple bands. The unique pixel binning scheme in \pixedfit, which takes into account the similarity in SED shape among pixels, allows for achieving a sufficient S/N ratio in multiple filters of interest while preserving important spatial information at the pixel level. A detailed description of this pixel binning scheme is given in \citet{2021Abdurrouf}. 

We assume the following parameters in the pixel binning process. We refer the reader to \citet{2021Abdurrouf} for more information about the parameters. We set S/N thresholds to $5$ in all \JWST\ NIRCam filters. We do not set the S/N threshold to \HST\ filters because the S/N ratio of pixels in the \HST\ images is low, especially for galaxies at high redshifts. Setting an S/N threshold on the \HST\ filters would put a strong constraint in the pixel binning process which can produce a coarser binning map and loosing important spatial information from the original images.

\begin{figure*}
\centering
\includegraphics[width=0.62\textwidth]{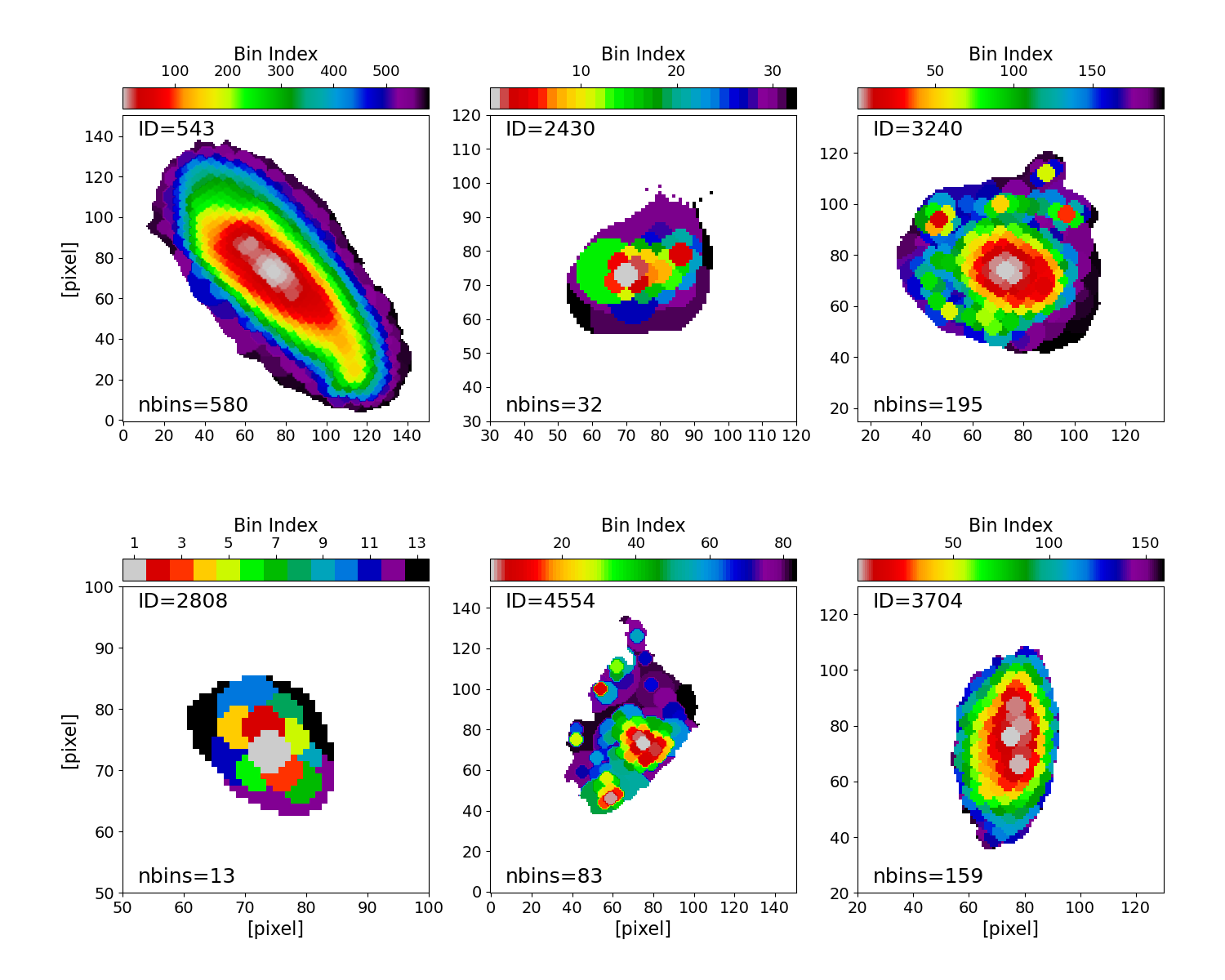}
\caption{Examples of pixel binning results. The pixel binning process achieves a minimum S/N ratio of 5 in all \JWST\ NIRCam filters.}
\label{fig:pixel_binning}
\end{figure*}

The rest of the binning parameters are as follows. We set a minimum diameter of 7 pixels, which is larger than the PSF FWHM size of our data cubes, a reduced $\chi^{2}$ limit of 5 in the evaluation of the similarity of SED shape. We refer to F277W flux in determining the brightest pixel to be the center of a spatial bin. We store new data cubes produced from this pixel binning process into FITS files. The total number of spatial bins in our sample galaxies is 24999. Figure~\ref{fig:pixel_binning} shows examples of the pixel binning maps produced from this process.

\subsection{Spatially Resolved SED Fitting} \label{sec:resolved_sedfit}

\begin{deluxetable*}{lp{8cm}p{5cm}l}
\tablecaption{Free Parameters in the SED Modeling and the Assumed Priors. \label{tab:params_sedfit}}
\tablewidth{0pt}
\tablehead{
\colhead{Parameter} & \colhead{Description} & \colhead{Prior} & \colhead{Sampling/Scale}
}
\startdata
$M_{*}$ & Stellar mass & Uniform: min$=\log(s_{\rm best})-2$, max$=\log(s_{\rm best})+2$\textsuperscript{a} & Logarithmic \\
$Z_{*}$ & Stellar metallicity & Uniform: min$=-2.0+\log(Z_{\odot})$, max$=0.2+\log(Z_{\odot})$ & Logarithmic \\
$t$ & Time since the onset of star formation ($\text{age}_{\rm sys}$) \textsuperscript{b} & Uniform: min$=-1.0$, max = age of the universe at the galaxy's redshift & Logarithmic \\
$\tau$ & Parameter that controls the peak time in the double power-law SFH model\textsuperscript{b} & Uniform: min$=-1.5$, max$=1.14$ & Logarithmic \\
$\alpha$ & Parameter in the double power-law SFH model that controls the slope of the falling star formation episode\textsuperscript{b} & Uniform: min$=-2.0$, max$=2.0$ & Logarithmic \\
$\beta$ & Parameter in the double power-law SFH model that controls the slope of the rising star formation episode\textsuperscript{b} & Uniform: min$=-2.0$, max$=2.0$ & Logarithmic \\
$\hat{\tau}_{1}$ & Dust optical depth of the birth cloud in the \citet{2000Charlot} dust attenuation law & Uniform: min$=0.0$, max$=4.0$ & Linear \\
$\hat{\tau}_{2}$ & Dust optical depth of the diffuse ISM in the \citet{2000Charlot} dust attenuation law & Uniform: min$=0.0$, max$=4.0$ & Linear \\
$n$ & Power law index in the \citet{2000Charlot} dust attenuation law & Uniform: min$=-2.2$, max$=0.4$ & Linear
\enddata
\tablecomments{
\textsuperscript{a}$s_{\rm best}$ is the normalization of model SED derived from the initial fitting with the $\chi^{2}$ minimization method (see Section 4.2.1 in \citealt{2021Abdurrouf}). 
\textsuperscript{b}The mathematical form of the double power-law SFH is given in \citet[][Equation 7 therein]{2021Abdurrouf}.
}   
\end{deluxetable*}
\vspace*{-2em}

Once we have the binned data cubes, we perform SED fitting to the SEDs of individual spatial bins in our sample galaxies. Here we use the SED fitting module in \pixedfit. The SED fitting in \pixedfit\ uses a fully Bayesian technique. We refer the reader to \citet{2021Abdurrouf} for a detailed description of the SED modeling and fitting methods as well as comprehensive tests of its capabilities. In Appendix~\ref{sec:fit_tests}, we perform SED-fitting tests using mock SEDs to demonstrate the robustness of our SED fitting method on combined \HST\ and \JWST\ photometry. Moreover, in Appendix~\ref{sec:breaking_degeneracies} we discuss how NIRCam photometry can potentially help in breaking the degeneracies among age, dust, and metallicity in SED fitting. In the following, we provide a brief description of the method and some assumptions applied in our SED fitting. 

We use the Flexible Stellar Population Synthesis code \citep[\texttt{FSPS};][]{2009Conroy,2010Conroy}. It includes the nebular emission modeling that uses the \texttt{CLOUDY} code \citep{1998Ferland, 2013Ferland}. In this work, we assume the \citet{2003Chabrier} initial mass function (IMF), Padova isochrones \citep{2000Girardi,2007Marigo,2008Marigo}, and MILES stellar spectral library \citep{2006Sanchez-Blazquez,2011Falcon}. For the star formation history model, we assume an analytic model in the form of a double power-law. It has been shown in \citet{2021Abdurrouf} that this SFH form can give robust inferences of the stellar population properties and even SFH of galaxies, as tested using synthetic SEDs of simulated galaxies in the IllustrisTNG simulations. For simulating the effect of dust attenuation, we use the two-component dust attenuation law of \citet{2000Charlot}. This dust attenuation law gives an extra attenuation to stars younger than $10$ Myr, in addition to standard attenuation in the diffuse ISM. We model the attenuation due to the intergalactic medium using \citet{2014Inoue} model. Since we do not have photometric data that covers the rest-frames mid-infrared (MIR) and far-infrared (FIR), we switch off the modeling of dust emission and AGN dusty torus emission in the analysis throughout this work. The SED modeling has 9 free parameters. We summarize these parameters along with the assumed priors in Table~\ref{tab:params_sedfit}. We assume a constant ionization parameter ($U$) of $0.01$ in the modeling of the nebular emission.

In the current analysis, we rely on photometric redshift for all of our sample galaxies because we do not have spectroscopic observations at the moment we carry out this analysis. To get redshift estimates of the galaxies, we perform SED fitting with \pixedfit\ in which redshift is let to be free in the fitting. For this, we fit integrated SED within the effective radius of the galaxies. The effective radius is measured in F444W image stamp using \galfit (\citealt{2002Peng}; see Section~\ref{sec:radial_profiles}). This is performed to get SEDs with high S/N while reducing contamination from noisy SEDs of pixels in the outskirt regions. In this fitting, we apply a prior on redshift in the form of a Gaussian function centered at the photometric redshift estimated by the \eazypy\ taken from the \grizli\ catalog (see Section~\ref{sec:data_reduction}). We set a width of $0.5$ for this Gaussian prior. This fitting is performed to derive redshift only. We then use this redshift information for the SED fitting of all spatial bins in the galaxy, in which we fix the redshift. We apply the Markov Chain Monte Carlo (MCMC) method in \pixedfit. In the SED fitting for redshift determination, we set the number of walkers to 100 and the number of steps per walker to 1000. For the SED fitting of spatial bins, we use 100 walkers and less number of steps per walker (600) for reducing computational time.

\begin{figure*}
\centering
\includegraphics[width=0.95\textwidth]{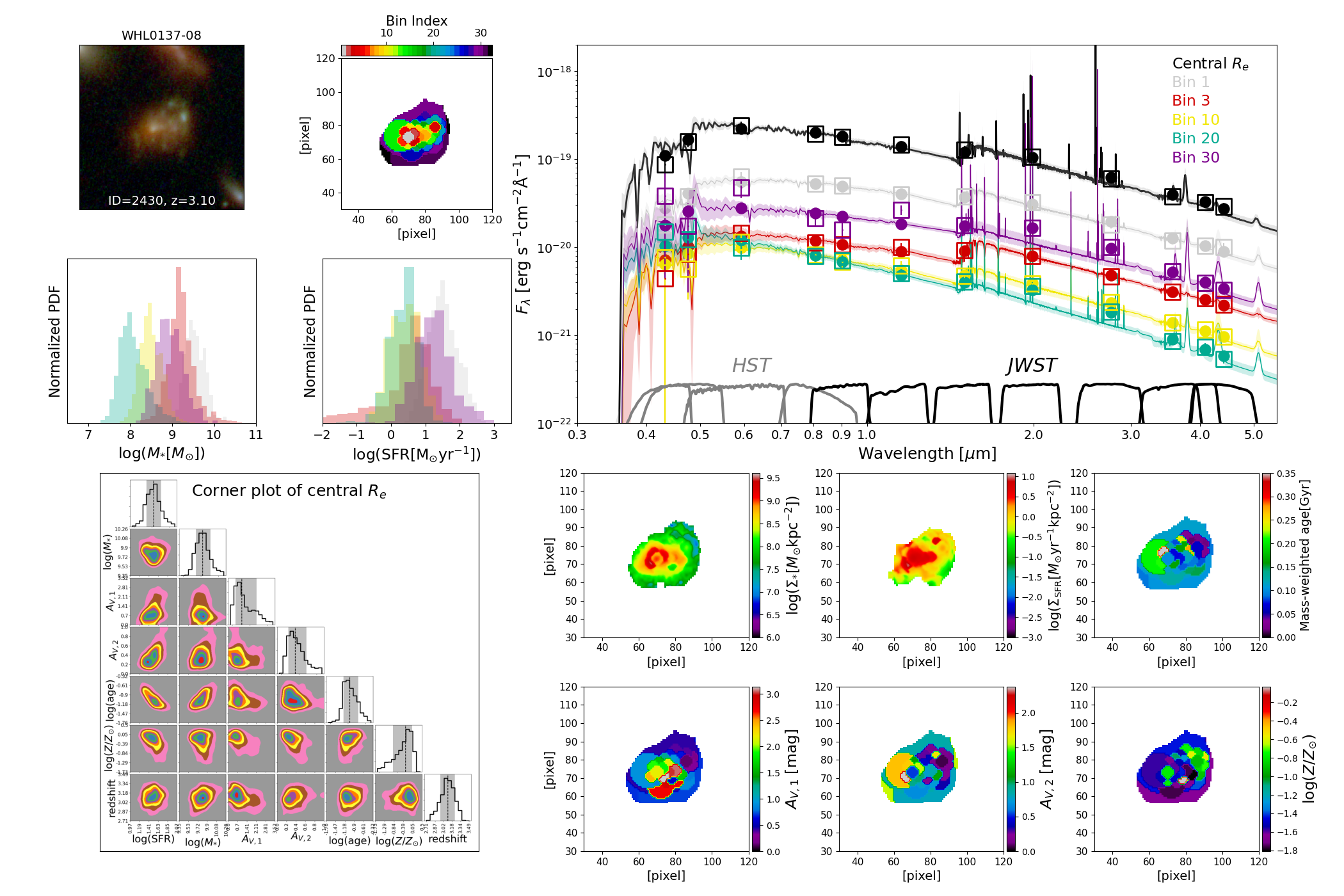}
\includegraphics[width=0.95\textwidth]{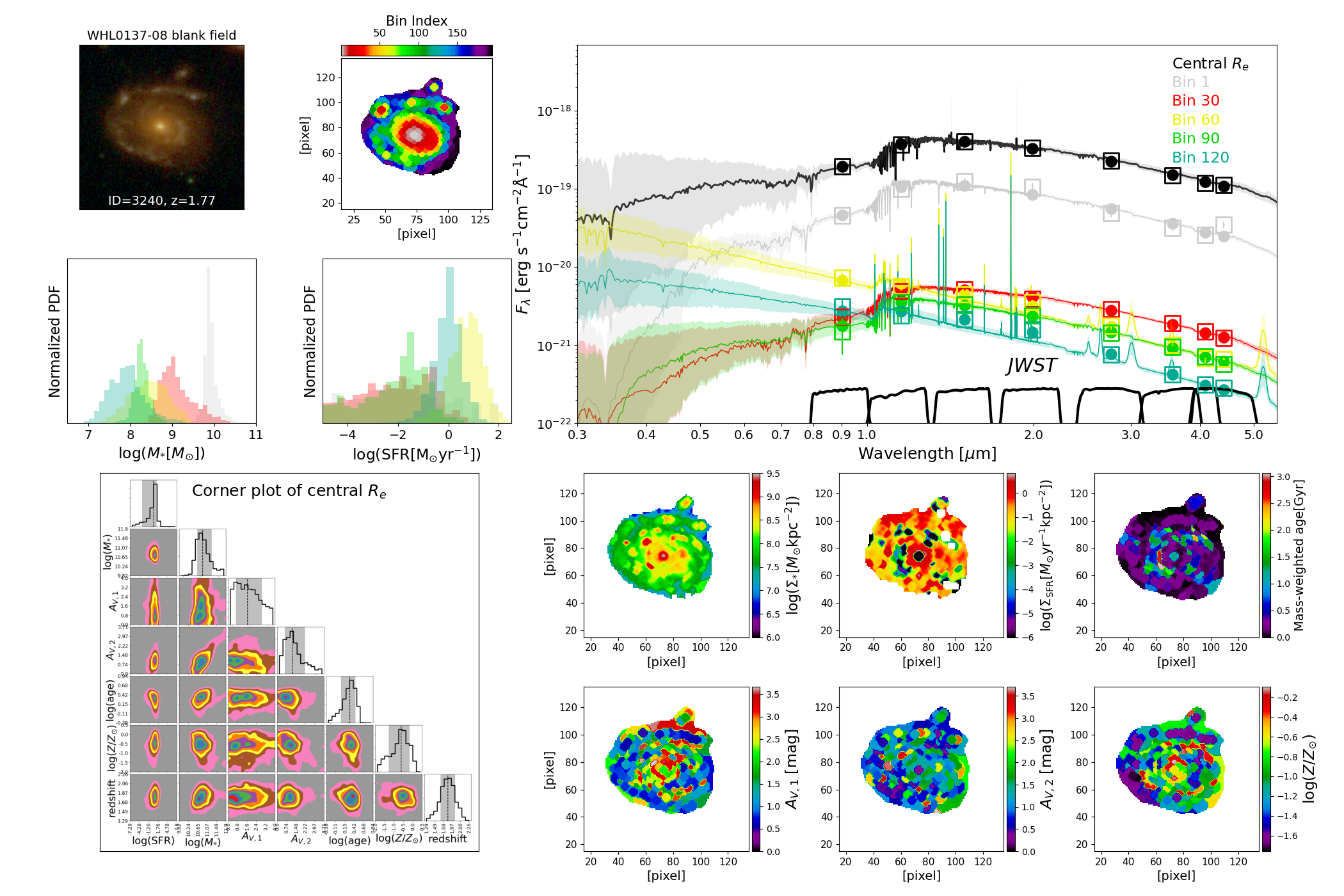}
\caption{Examples of SED fits of a galaxy in the \WHL\ cluster (top panel) and the blank field (bottom panel). The SED plots show the best-fit SEDs from the fitting to integrated SED within the effective radius (black color) and 5 examples of spatial bins (in colors). The corner plots show the posterior probability distributions of the parameters obtained from fitting to the central integrated SED. Above this corner plot, we show PDFs of \mass\ and SFR of the 5 example spatial bins. Finally, the maps of stellar population properties derived from this analysis are shown in the bottom right corner.}
\label{fig:plot_sedfits}
\end{figure*}

We show examples of SED fitting results of two galaxies in Figure~\ref{fig:plot_sedfits}, one galaxy from the \WHL\ cluster field (top panel) and the other galaxy from the blank field (bottom panel). For each galaxy, we show best-fit SEDs in the top right panel. The observed and best-fit photometric SEDs are shown with square and circle symbols, respectively. The SED in black represents the integrated SED within the effective radius, while those in other colors are for 5 examples of spatial bins in the galaxies. The corner plot in the bottom left side shows the posterior probability distribution functions (PPDF) of the model parameters obtained from the fitting on the integrated SED within the effective radius. Above this corner plot, we show the PDFs of \mass\ and SFR of the example spatial bins. The best-fit spectra shown in the plot are drawn from the MCMC sampler chains. Therefore, it is possible to get a slight shift in wavelength between the best-fit spectra of the central SED (where $z$ is free in the fitting) and that of the spatial bins (where $z$ is fixed in the fitting). This wavelength shift reflects the uncertainty of the estimated redshift. Finally, in the bottom right panel, we show the maps of stellar population properties, including the $M_{*}$ surface density (\massd), SFR surface density (\sfrd), mass-weighted age, $A_{V,1}$ ($\equiv 1.086\times \hat{\tau}_{1}$), $A_{V,2}$ ($\equiv 1.086\times \hat{\tau}_{2}$), and metallicity.               

\subsection{Lens Modeling} \label{sec:lens_models}

To estimate the magnifications due to the gravitational lensing effect by the clusters, we use the lens models constructed by our team. For the \WHL\ cluster, we use the same lens models that were used for analyzing the Earendel and the Sunrise Arc in \citet{2022Welch1}, which were made publicly available\footnote{\url{https://relics.stsci.edu/lens_models/outgoing/whl0137-08/}}. These lens models were generated using four independent lens modeling software packages: \texttt{Light-Traces-Mass} \citep[\texttt{LTM},][]{Zitrin09, Zitrin15, Broadhurst05}, \texttt{Glafic} \citep{Oguri2010}, \texttt{WSLAP} \citep{Diego05wslap, Diego07wslap2}, and \texttt{Lenstool} \citep{1993Kneib, JulloLenstool07, JulloLenstool09}. Please refer to \citet{2022Welch1} for detailed information about each model. Sample galaxies located in the blank field are expected to have only weak magnifications of $\mu \lesssim 1.1$. With the multiple lens models available for this cluster, we estimate the total and tangential (i.e.,~linear) magnifications ($\mu$ and $\mu_{t}$, respectively) of each galaxy by taking the average values. In this way, we account for the modeling uncertainties. Based on the standard deviation values, we find that the magnifications do not vary a lot among the models. The median standard deviations of $\mu$ and $\mu_{t}$ are $0.11$ and $0.10$ dex, respectively.

The lens models for \MACS\ cluster have been constructed in the past using the \HST\ imaging data. The first lens model for \MACS, before the CLASH survey, was provided by \citet{2011Zitrin} using \texttt{LTM} method. With the addition of \HST\ imaging data from CLASH, new lens models were established using various methods, including \texttt{Lenstool}, \texttt{LTM}, \texttt{WSLAP}, and \texttt{LensPerfect} \citep{2008Coe}. These lens models have been used in previous studies in CLASH \citep[e.g.,][]{2013Coe, Zitrin15, 2017Chan}. Now with the addition of \JWST\ NIRCam imaging data, which add on many new strongly-lensed multiple-image candidates (thanks to its high spatial resolution and depth), a new lens model has been established using the \texttt{dPIEeNFW} method \citep{Zitrin15} with some modifications. Detailed information on this lens modeling of \MACS\ cluster along with the list of the multiple-image systems considered in the modeling are given in \citet{2022Meena}. This new method has also been implemented in several clusters using \JWST\ NIRCam data \citep{2022Pascale,2022Roberts-Borsani,2022Hsiao,2022Williams}. We used this lens model constructed by \citet{2022Meena} for galaxies in the \MACS\ field. We correct the $M_{*}$ and SFR obtained from SED fitting for the lensing magnification by dividing them with $\mu$. We also correct size or radius measurements by dividing them with $\mu_{t}$.

\section{Results} \label{sec:results}

\subsection{Integrated Properties} \label{sec:global_SFMS}

\begin{figure*}
\centering
\includegraphics[width=0.47\textwidth]{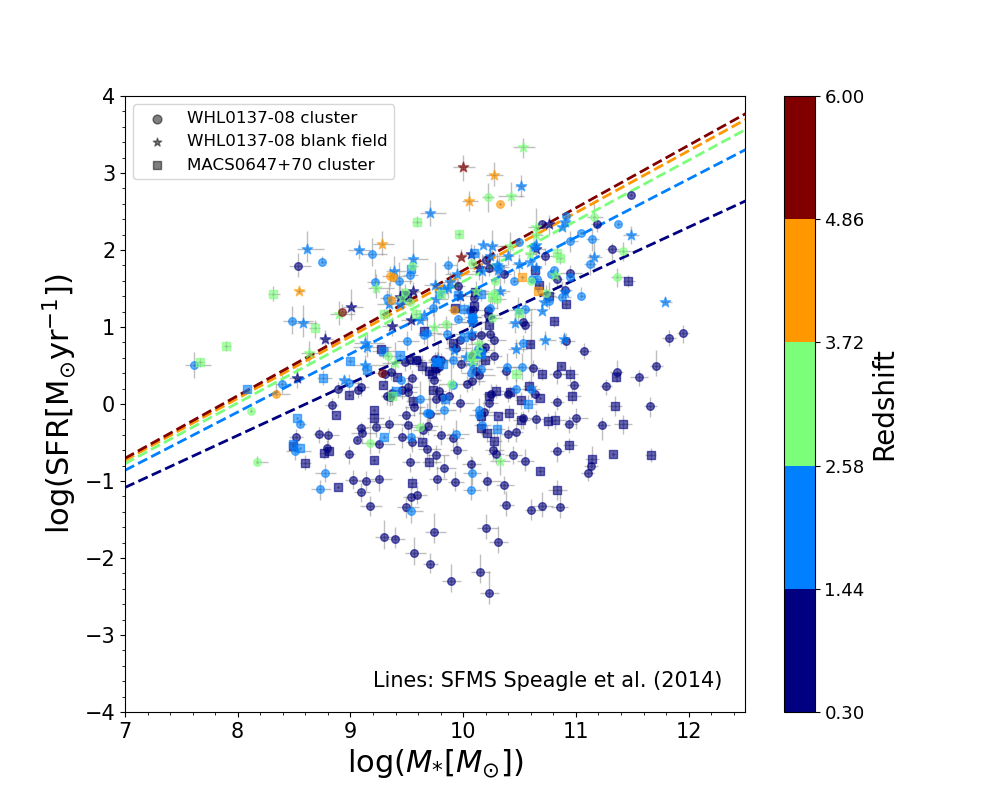}
\includegraphics[width=0.51\textwidth]{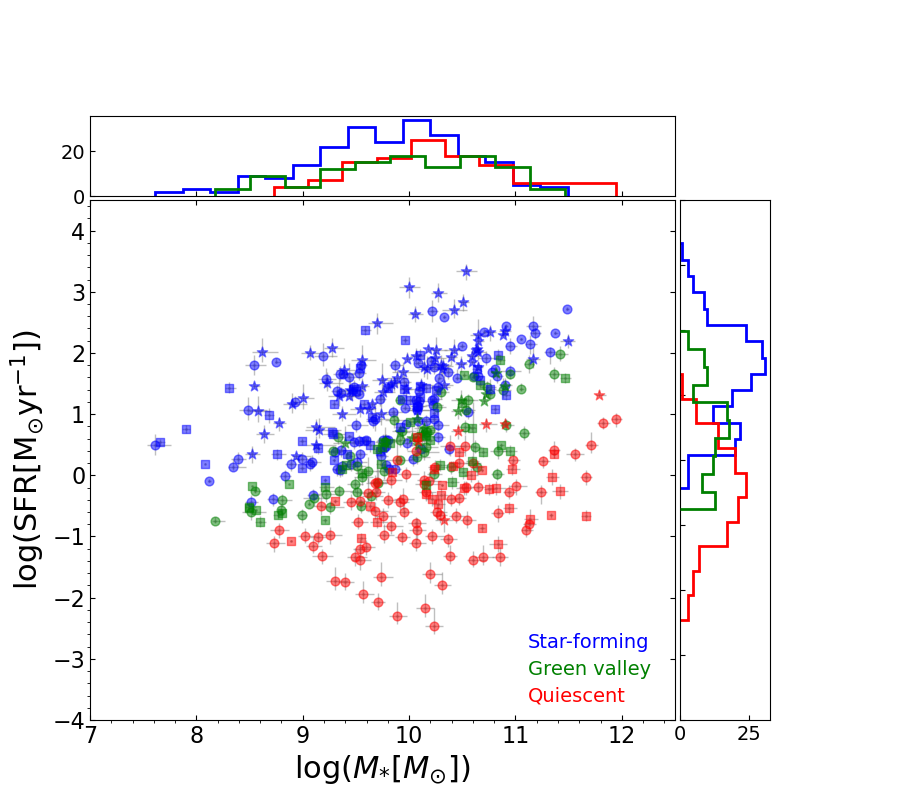}
\caption{\textit{Left panel}: Integrated (i.e.,~global) $M_{*}$ and SFR of our sample galaxies. Different symbols represent different fields where the galaxies are located, whereas the color-coding represents redshift grouping. The dashed lines show the global SFMS relations at the median redshifts of the 5 redshift groups calculated using the prescription from \citet{2014Speagle}. The lines are colored based on the redshift groups. \textit{Right panel}: Distribution of the star-forming, green-valley, and quiescent galaxies in our sample that are classified as having $\text{SFR}>\text{SFR}_{\rm MS}(z,M_{*})-0.4$ dex, $\text{SFR}_{\rm MS}(z,M_{*})-0.4\geq \text{SFR} > \text{SFR}_{\rm MS}(z,M_{*})-1.0$ dex, and $\text{SFR}\leq \text{SFR}_{\rm MS}(z,M_{*})-1.0$ dex, respectively. $\text{SFR}_{\rm MS}(z,M_{*})$ is the SFMS ridge line that is calculated for exact $z$ and $M_{*}$ of the individual galaxies.}
\label{fig:global_sfms}
\end{figure*}

Before analyzing the spatially resolved properties of our sample galaxies, we first present their integrated (i.e.,~global) properties. To bring it into the context of the global demographics of galaxies, we plot our sample on the integrated star-forming main sequence (SFMS) diagram, as shown in the left panel of Figure~\ref{fig:global_sfms}. The integrated $M_{*}$ and SFR of a galaxy are derived by summing up the values in pixels obtained from the spatially resolved SED fitting. Due to our limited sample, we plot all our galaxies on the SFMS diagram instead of dividing them into a number of redshift bins and examining the SFMS relation in each bin. This can cause a broad distribution as shown in the figure. Different symbols represent the fields where the galaxies are located (\WHL, blank field, and \MACS), whereas color-coding represents redshift grouping, where we divide the redshift range into five bins. The dashed lines show the SFMS relations at the median redshifts of the five redshift bins, calculated using the prescription from \citet{2014Speagle}. The lines are colored based on the redshift groups. 

We then classify our sample galaxies into star-forming, green valley, and quiescent groups based on their positions with respect to the SFMS ridge line at the redshift of the galaxies. We define star-forming, green-valley, and quiescent galaxies as those having $\text{SFR}>\text{SFR}_{\rm MS}(z,M_{*})-0.4$ dex, $\text{SFR}_{\rm MS}(z,M_{*})-0.4\geq \text{SFR} > \text{SFR}_{\rm MS}(z,M_{*})-1.0$ dex, and $\text{SFR}\leq \text{SFR}_{\rm MS}(z,M_{*})-1.0$ dex, respectively, where $\text{SFR}_{\rm MS}(z,M_{*})$ is the SFMS ridge line for exact $z$ and $M_{*}$ of the individual galaxies. With this selection criteria, we have 219, 108, and 117 total numbers of the star-forming, green-valley, and quiescent galaxies from the three fields, respectively. We will use these classified samples throughout the analysis in this paper to investigate the differences in spatially resolved properties of galaxies in various evolutionary stages. The right panel of Figure~\ref{fig:global_sfms} show the distributions of these galaxy groups on the SFMS diagram.   

\begin{figure}
\centering
\includegraphics[width=0.52\textwidth]{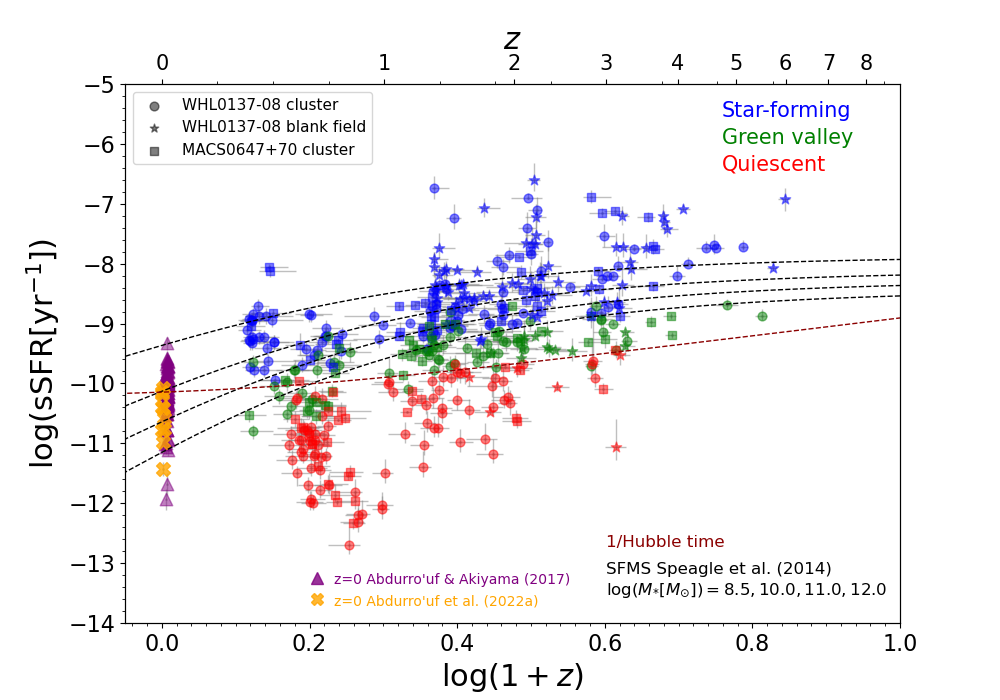}
\caption{Evolution of the integrated specific SFR (sSFR) with redshift. The black dashed lines represent sSFR evolution of SFMS galaxies with $M_{*}=10^{8.5}$, $10^{10}$, $10^{11}$, and $10^{12}$ $M_{\odot}$ (decreasing normalization) as inferred from the normalization of the SFMS, calculated using the prescription from \citet{2014Speagle}. The red dashed line represents $1/t_{H}$ where $t_{H}$ is the Hubble time. Almost all our quiescent galaxies lie below this line, indicating that their mass doubling timescale is longer than $t_{H}$.}
\label{fig:global_sSFR_vs_z}
\end{figure}

To get a sense of how the global specific SFR (sSFR$\equiv \text{SFR}/M_{*}$) evolves with cosmic time in our sample galaxies, we plot the sSFR against redshift in Figure~\ref{fig:global_sSFR_vs_z}. We can see a clear trend of decreasing global sSFR with cosmic time and an increasing number of quiescent galaxies along the way. In our sample, quiescent galaxies start to emerge from $z\sim 3$, around $2$ Gyr after the Big Bang. To compare our global sSFR trend with the similar trend from previous studies, we plot sSFR$(z)$ inferred from the SFMS normalization based on \citet{2014Speagle} prescription. We calculate sSFR$(z)$ with four $M_{*}$ of $10^{8.5}$, $10^{10}$, $10^{11}$, and $10^{12}$ $M_{\odot}$ and show them in the figure as black dashed lines. We see an overall agreement between the evolutionary trend of sSFR in our sample and that expected based on the evolution of the SFMS normalization. We also show the global sSFR of local ($z\sim 0$) galaxies from \citet{2017Abdurrouf} and \citet{2022Abdurrouf1} (personal communication), which were derived from spatially resolved SED fitting. \citet{2017Abdurrouf} analyzed 93 spiral galaxies at $0.01<z<0.02$ using imaging data from the GALEX \citep{2007Morrissey} and SDSS \citep{2000York} surveys. \citet{2022Abdurrouf1} applied \pixedfit\ for analyzing 10 nearby galaxies using imaging data in more than 20 filters spanning from Far-ultraviolet (FUV) to FIR. 

Previous studies have classified passive galaxies using various methods. One of the methods is by comparing the Hubble time ($t_{H}$) with the mass doubling time (i.e.,~inverse of sSFR). Basically, this method defines quiescent galaxies as those having $\text{sSFR}<1/t_{H}$. The red dashed line in Figure~\ref{fig:global_sSFR_vs_z} represents $1/t_{H}$. We can see that our quiescent galaxies lie below this line, indicating that our classification method is consistent with that based on $t_{H}$. 

\subsection{Radial Profiles of the Stellar Population Properties} \label{sec:radial_profiles}

As we have shown the global properties of the sample galaxies and classified them into star-forming, green-valley, and quiescent groups, now we will analyze their spatially resolved properties. We start by presenting the radial profiles of the stellar population properties to get a sense of how the properties vary radially within the galaxies. To derive the radial profiles, first, we perform 2D single-component S\'ersic fitting using \galfit\ \citep{2002Peng} on F444W stamp image of each galaxy to get their ellipticities, position angles, and central coordinates. We then use this information to define elliptical annuli in the radial profile calculation. The radial profiles are derived from the 2D maps of properties obtained from the spatially resolved SED fitting by averaging values of pixels within the annuli. Since galaxies have a wide range of size, we normalize the radius by the half-mass radius ($R_{e}$), which is the radius that covers half of the integrated $M_{*}$. We use radial increment ($\delta r$) of $0.3R_{e}$. Thanks to the gravitational lensing effect, we can resolve many of our galaxies down to sub-kpc scales (109 galaxies in our sample have delensed $R_{e}<1$ kpc).    

\begin{figure*}
\centering
\includegraphics[width=0.85\textwidth]{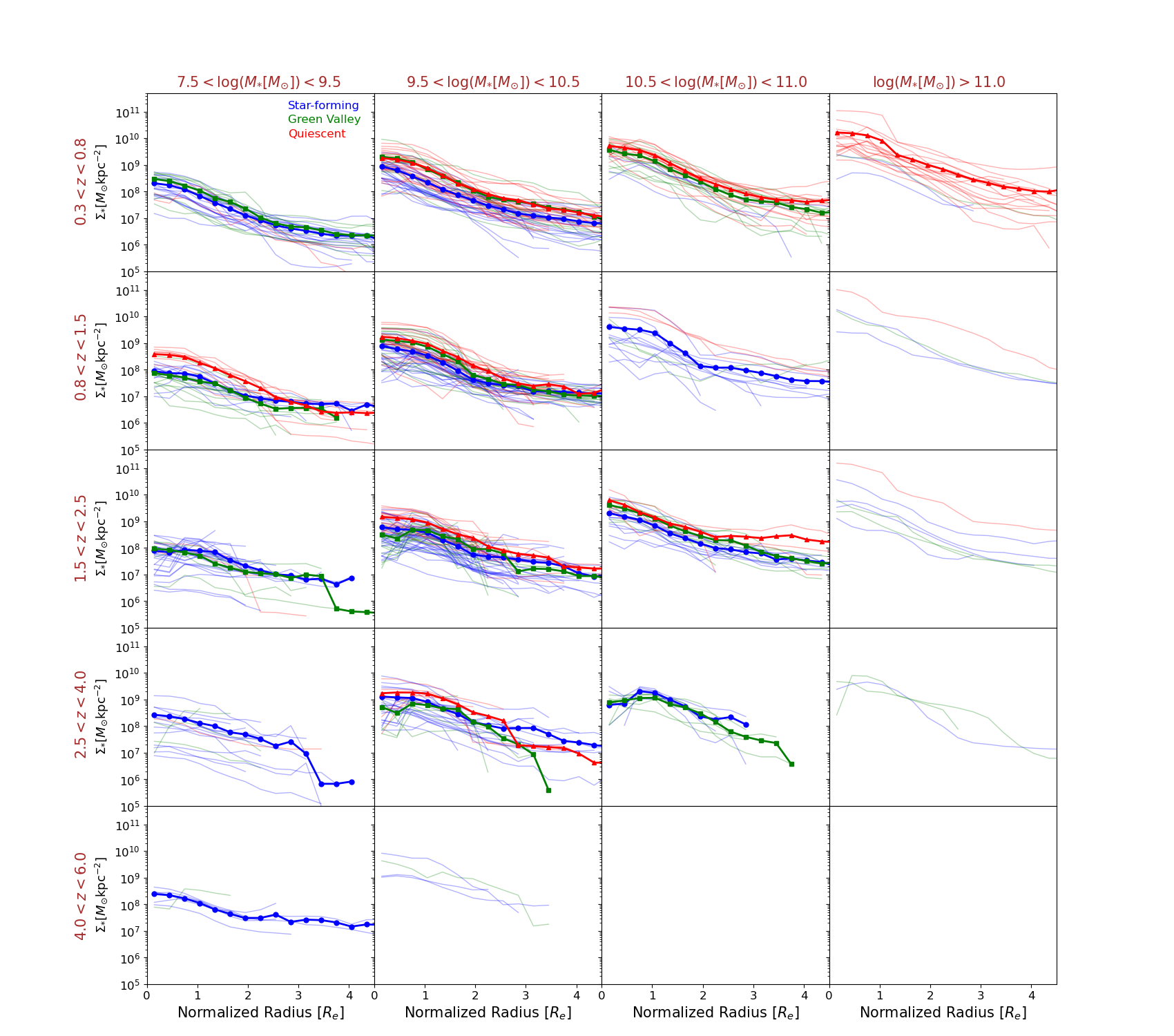}
\caption{Radial profiles of the stellar mass surface density ($\Sigma_{*}$). The sample galaxies are divided into 4 bins of global $M_{*}$ and 5 bins of redshift. At each group, we further classify the galaxies into star-forming, green-valley, and quiescent groups and indicate them with different colors. For sub-groups that contain at least 5 galaxies, we show average radial profiles with tick lines. At each redshift bin, more massive galaxies tend to have higher $\Sigma_{*}(r)$ normalization than less massive galaxies, indicating that the excess in mass happens across the galaxy region. Quiescent galaxies tend to have higher $\Sigma_{*}(r)$ normalization than star-forming in all redshifts. This is especially clear in high $M_{*}$ bins.}
\label{fig:radial_profiles_mass}
\end{figure*}

Figure~\ref{fig:radial_profiles_mass} shows the radial profiles of the stellar mass surface density ($\Sigma_{*}$). We divide the sample into 5 bins of redshift and 4 bins of $M_{*}$ to see how the radial profiles vary with global $M_{*}$ and cosmic time. Moreover, we indicate the star-forming, green valley, and quiescent galaxies with different colors, in a similar way as in Figure~\ref{fig:global_sSFR_vs_z}. For groups  that contain at least 5 galaxies, we show average radial profiles with tick lines. Some interesting trends from Figure~\ref{fig:radial_profiles_mass} are the following. At each redshift bin, more massive galaxies tend to have higher normalization of $\Sigma_{*}(r)$ than less massive ones, indicating that the excess in mass happens across the entire radius. Moreover, we also see that quiescent galaxies tend to have higher $\Sigma_{*}(r)$ normalization than the star-forming and green-valley galaxies in all redshift. This is especially clear in the most massive groups. It is also interesting to see that $\Sigma_{*}(r)$ profiles have a negative gradient (i.e.,~decreasing mass with increasing radius) in all redshift and mass bins, although the profiles seem to be shallower at higher redshifts. 

\begin{figure*}
\centering
\includegraphics[width=0.85\textwidth]{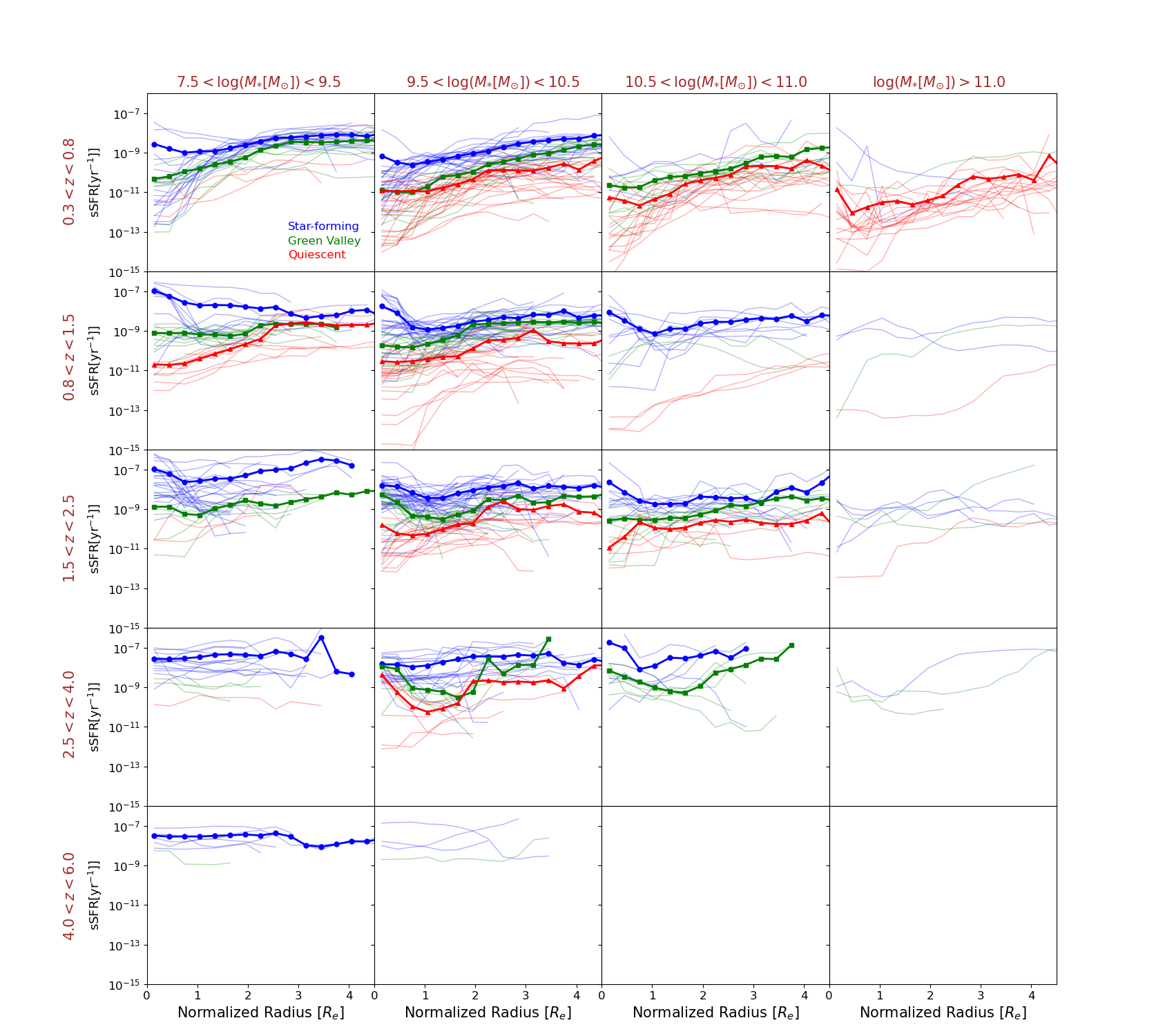}
\caption{Similar to Figure~\ref{fig:radial_profiles_mass} but for radial profiles of sSFR. At $z\gtrsim 2.5$, our sample galaxies tend to have a broadly flat sSFR$(r)$, while at the lower redshifts, they show more diversity in shape. At $0.8\lesssim z\lesssim 2.5$, the majority of star-forming galaxies have an elevation of sSFR in their central regions, while they tend to have centrally-suppressed sSFR radial profiles at lower redshifts. On the other hand, quiescent galaxies in our sample tend to have centrally-suppressed sSFR radial profiles since $z\sim 1.5$.}
\label{fig:radial_profiles_sSFR}
\end{figure*}

To see how galaxies quench their star formation, specifically where in the galaxies the suppression of star formation first happens and how it progresses over cosmic time, next we analyze the radial profiles of sSFR. The radial profiles of sSFR are shown in Figure~\ref{fig:radial_profiles_sSFR}. As we can see from this figure, the sSFR radial profiles of the majority of our sample galaxies at $z\gtrsim 2.5$ are broadly flat, while they show more diversity in shape at lower redshifts. At $0.8\lesssim z\lesssim 2.5$, star-forming galaxies in our sample tend to have a flat or centrally-peaked sSFR$(r)$, while quiescent galaxies tend to have centrally-suppressed sSFR$(r)$. On the other hand, green-valley galaxies in our sample seem to have broadly flat radial profiles up to $z\sim 1.0$, except in the most massive group, where some of them show an sSFR suppression in their central regions. At lower redshifts, the majority of our sample galaxies have centrally-suppressed sSFR$(r)$. It is also interesting to see that the majority of star-forming galaxies at $0.8\lesssim z\lesssim 2.5$ (in which the cosmic noon epoch is covered), have a centrally-peaked sSFR$(r)$. This central elevation of sSFR is not observed at higher redshifts.       

\begin{figure*}
\centering
\includegraphics[width=0.85\textwidth]{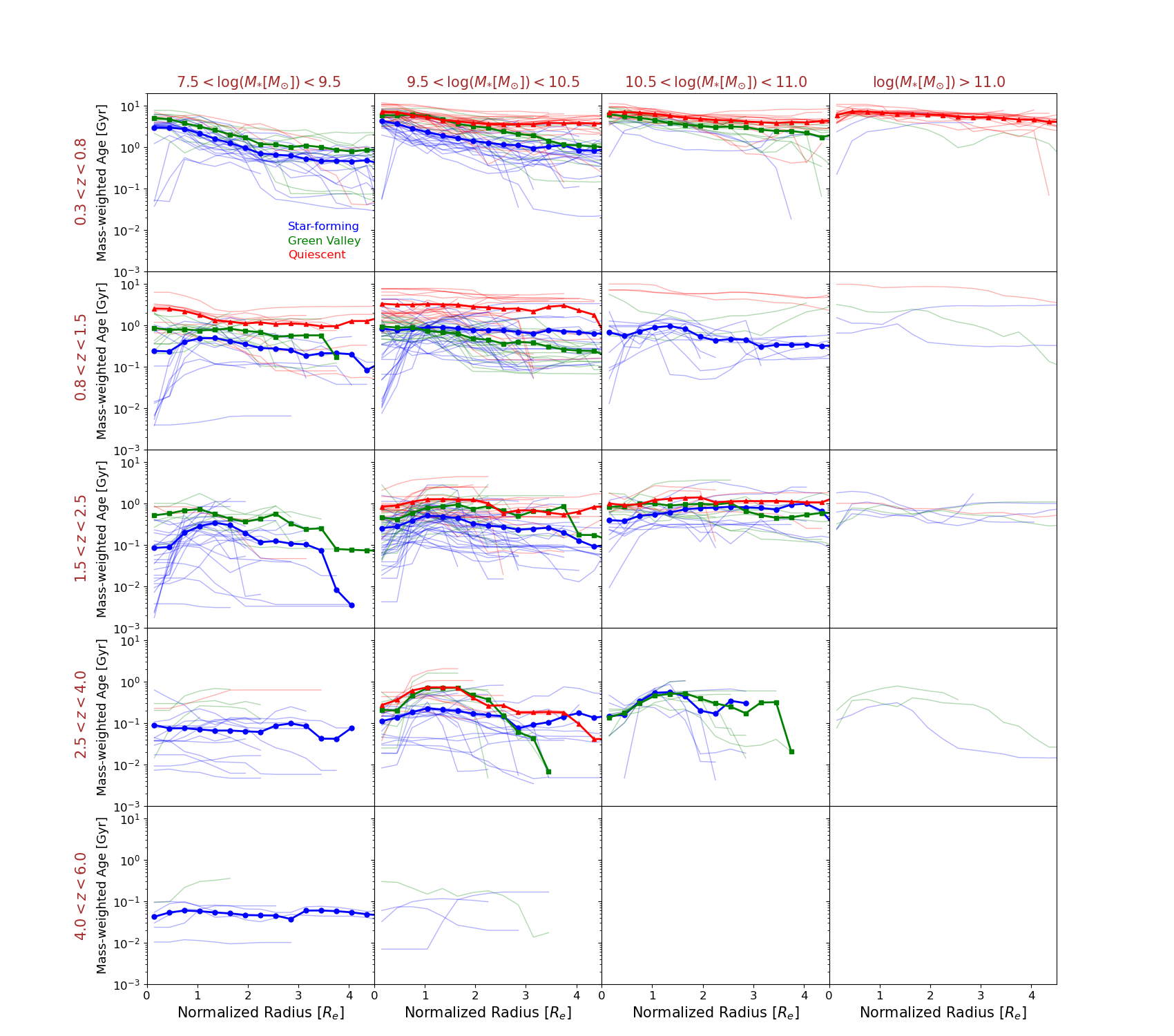}
\caption{Similar to Figure~\ref{fig:radial_profiles_mass} but for radial profiles of the stellar population age. There is a trend of increasing the overall age of the stellar populations in galaxies over cosmic time. At $0.8\lesssim z\lesssim 2.5$ (possibly around the cosmic noon epoch), a large fraction of our star-forming galaxies have a centrally-suppressed age radial profile. Those galaxies have centrally-peaked sSFR radial profiles as shown in Figure~\ref{fig:radial_profiles_sSFR}. These trends indicate that they are likely to have an ongoing nuclear starburst event.}
\label{fig:radial_profiles_age}
\end{figure*}

Next, we analyze the radial profiles of the stellar population age to see how this quantity varies radially within our sample galaxies and investigate the underlying stellar population properties causing the diversity in the sSFR radial profiles. From our spatially resolved SED fitting, we obtain maps of the mass-weighted ages, which is the average age of stars in a stellar population as weighted by the stellar mass formed over the course of the star formation history. The age radial profiles are shown in Figure~\ref{fig:radial_profiles_age}. As can be seen from this figure, there is a trend of increasing overall age of the stellar populations in galaxies over cosmic time, as indicated by the increasing normalization of the radial profiles with decreasing redshift. The star-forming galaxies that have a centrally-peaked sSFR$(r)$ at $0.8\lesssim z\lesssim 2.5$ (possibly around the cosmic noon epoch) as shown in Figure~\ref{fig:radial_profiles_sSFR} are likely in a phase of rapid star formation in their centers (i.e.,~a nuclear starburst; e.g.,~\citealt{2014Dekel,2015Zolotov,2016Tacchella,2017Tadaki}), as indicated by the young stellar populations (age $\lesssim 100$ Myr) in their central regions. At this epoch, green-valley and quiescent galaxies tend to have radially decreasing age profiles (i.e.,~negative gradient). At $0.3<z<0.8$, low-mass galaxies ($\log(M_{*}/M_{\odot})<9.5$) in all stages of star formation have radially decreasing age radial profiles (i.e.,~negative gradient). A similar trend still holds for star-forming and green-valley galaxies in the higher mass group ($9.5<\log(M_{*}/M_{\odot})<10.5$). On the other hand, quiescent galaxies at this epoch tend to have overall flat and old stellar populations across their entire radius, with higher normalization (i.e.,~older) than that of star-forming and green-valley galaxies.

\subsection{Compactness of the Spatial Distributions of Stellar Mass and SFR} \label{sec:comp_mass_vs_SFR}

The centrally-peaked sSFR$(r)$ of star-forming galaxies at around the cosmic noon epoch indicates that they are likely undergoing a nuclear starburst that builds the bulge component, as has also been observed by previous studies \citep[e.g.][]{2017Tadaki,2022Kalita}. The centrally-suppressed sSFR$(r)$ profiles which start to emerge in quiescent galaxies at around the same epoch can be caused by the cessation of star formation in the center and/or a matured bulge that has been formed in these galaxies. This trend provides a hint on how galaxies quench their star formation, which seems to progress in an inside-to-outside manner (i.e.,~quenching starts from the center and then propagates outward). At the same time, this trend may indicate that galaxies build their central regions first, forming a mature bulge, and then subsequently assemble their disk through star formation (i.e.,~inside-out growth). To further investigate this, next we compare the compactness of the spatial distributions of \mass\ and SFR by means of the half-mass and half-SFR radii.

\begin{figure}[ht]
\centering
\includegraphics[width=0.45\textwidth]{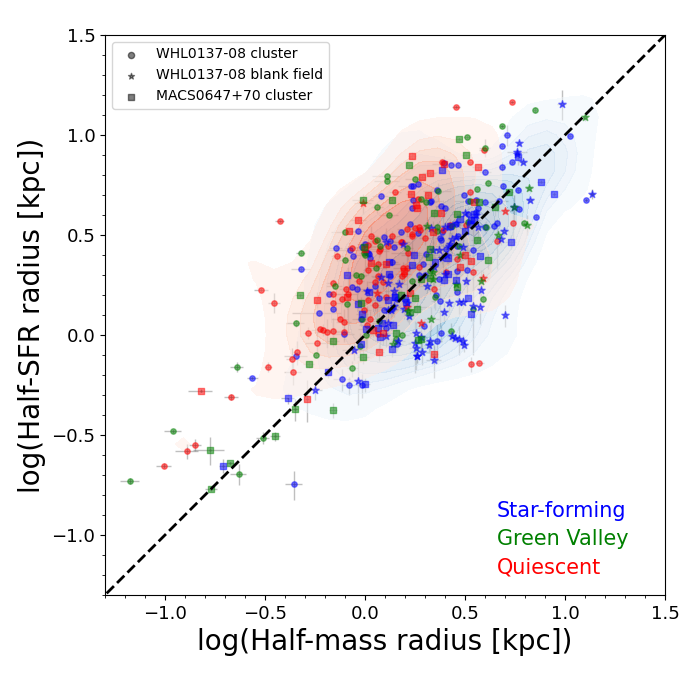}
\caption{Comparison between the half-mass radius and half-SFR radius. The overall symbols in this figure are the same as those in Figure~\ref{fig:global_sSFR_vs_z}. The red and blue contours show the number densities of the star-forming and quiescent galaxies. The majority of quiescent galaxies in our sample have spatially more extended star formation distribution than the stellar mass. Star-forming galaxies are equally distributed on the diagram. In star-forming galaxies with compact SFR distribution (i.e.,~below the one-to-one line), an active star formation happens in the central region. On the other hand, in star-forming galaxies with less compact SFR distribution, a bulge might have been built and active star formation is now progressing outward, building the disk.}
\label{fig:effradSM_vs_effradSFR}
\end{figure}

We compare the half-mass radius and the half-SFR radius in Figure~\ref{fig:effradSM_vs_effradSFR}. The half-SFR radius is a radius (measured along the elliptical semi-major axis) that covers half of the total SFR. To compare the distributions of our star-forming and quiescent galaxies on this diagram, we plot the density contours. As can be seen from this figure, star-forming galaxies broadly follow the one-to-one line, whereas quiescent galaxies are in excess above the line. This means that in quiescent galaxies, the spatial distribution of SFR is more extended than that of stellar mass, indicating that star formation is ongoing in the disk and less active in the central region. It is also possible that a massive bulge might have been formed in the centers, making a more compact stellar mass distribution. On the other hand, star-forming galaxies are equally distributed. Some star-forming galaxies have spatially more compact star formation distribution than the stellar mass (i.e.,~below the one-to-one line), which indicates that active star formation happens at their centers. On the other hand, in the star-forming galaxies that have extended star formation (i.e.,~above the one-to-one line), the bulge might have been built and active star formation is now progressing outward and building the disk.

It has been known that galaxy size correlates with global $M_{*}$ for galaxies out to at least $z\sim 3$ \citep[i.e.,~the size--mass relation; e.g.,~][]{2003Shen, 2014vanderWel, 2014morishita, 2021Yang}. However, most of the previous studies rely on galaxy half-light radii as a measure of galaxy size. Since mass-to-light ratios are not constant across a galaxy's region but instead have a gradient, the half-light radii are not a direct probe of the underlying stellar mass profiles. Therefore, it is expected that the half-mass and half-light radii are different. The difference in size as probed by light and mass profiles in galaxies out to $z\sim 2.5$ has been investigated by previous studies \citep[e.g.,][]{2019Suess}. 

\begin{figure}[ht]
\centering
\includegraphics[width=0.45\textwidth]{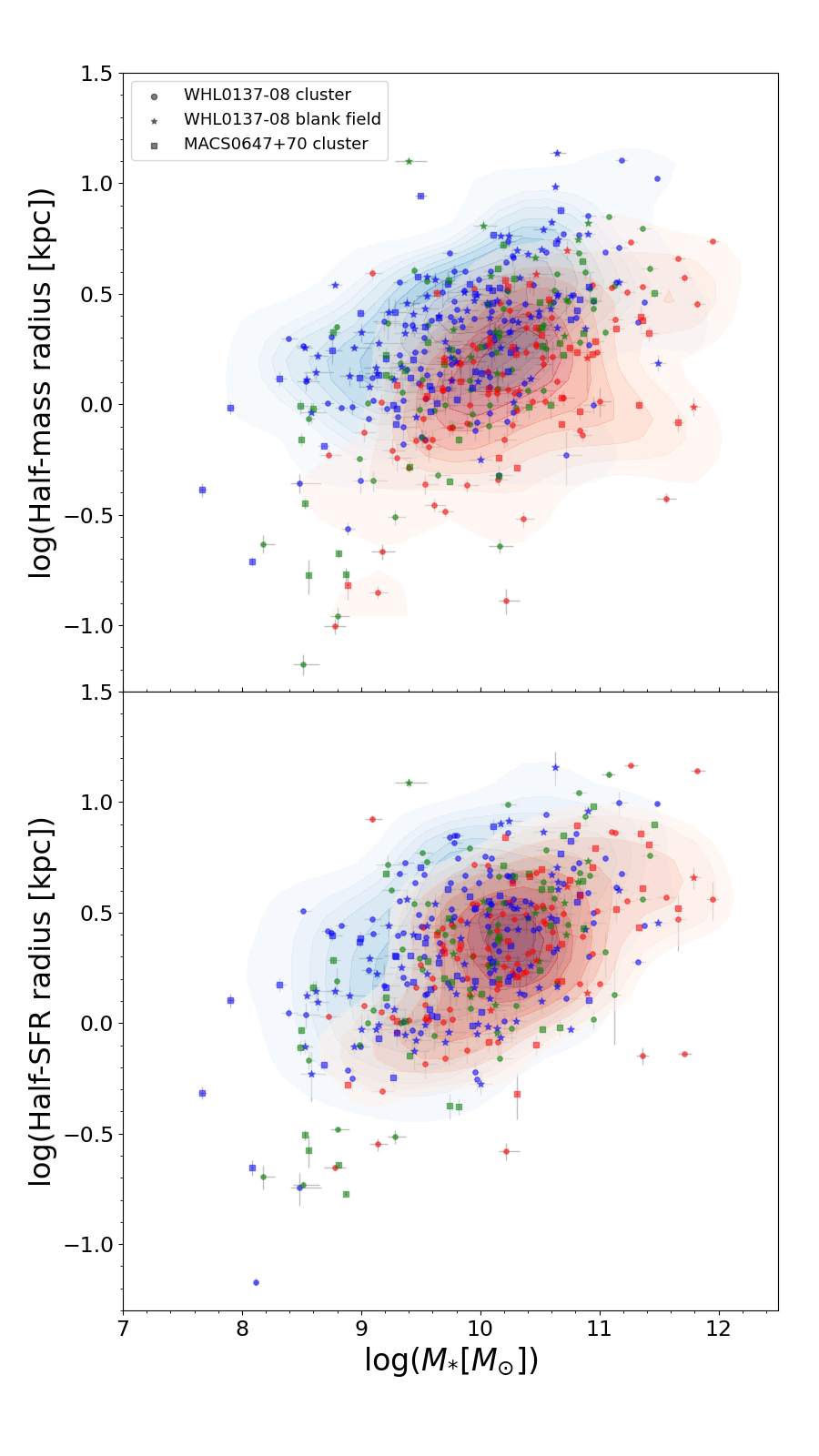}
\caption{The half-mass (top panel) and half-SFR (bottom panel) radii as a function of integrated \mass. The overall symbols in this figure are the same as those in Figure~\ref{fig:effradSM_vs_effradSFR}. The half-mass radius is defined as the radius that covers half of the total \mass, while the half-SFR radius is the radius covering half of the total SFR. Both half-mass and half-SFR radii increase with increasing \mass. The \mass\ vs half-mass radius diagram shows that overall, the star-forming galaxies tend to have a larger size than quiescent galaxies. However, there is no clear difference between the two groups in terms of the size of the SFR distribution.}
\label{fig:mass_vs_effrad_SM}
\end{figure}

Now, we check how half-mass and half-SFR radii correlate with integrated \mass\ in our sample galaxies. We show the correlations in Figure~\ref{fig:mass_vs_effrad_SM}. Overall, we see that both half-mass and half-SFR radii increase as increasing integrated \mass. However, we observe a difference in how star-forming and quiescent galaxies are distributed in the two correlations. In the top panel, we can see that star-forming galaxies tend to have a larger half-mass radius than quiescent galaxies in all masses. On the other hand, as can be seen from the bottom panel, there is no clear difference between the star-forming and quiescent galaxies in terms of the half-SFR radius, although star-forming galaxies tend to have a wider range of half-SFR radius than quiescent galaxies. 

In agreement with the trend we observe here, previous studies using half-light radii also found that star-forming galaxies are larger than quiescent galaxies in all \mass\ \citep{2014vanderWel}. However, when the half-mass radius is used, the normalization of the relation decreases, and the slope as well becomes shallower \citep{2019Suess}. We do not intend to measure the slope and normalization of our size--mass relations because of the limited sample. Ideally, we need a large sample of galaxies and divide them into some redshift bins. In this work, we combine all galaxies in our sample despite the fact that they are located in a wide range of redshift.    

\subsection{Central Stellar Mass Surface Density}\label{sec:cent_mass_sd}

We have observed that quiescent galaxies have a smaller half-mass radius than star-forming galaxies. The quiescent galaxies at $z\lesssim 2.5$ also have centrally-suppressed sSFR radial profiles. The compactness of quiescent galaxies can be caused by the massive bulge that has been formed in their centers. This massive bulge may also be the reason behind the central suppression of sSFR. Therefore, here we investigate the stellar mass surface density in the central $1$ kpc radius (\sigmakpc) of our sample galaxies. 

\begin{figure}[ht]
\centering
\includegraphics[width=0.45\textwidth]{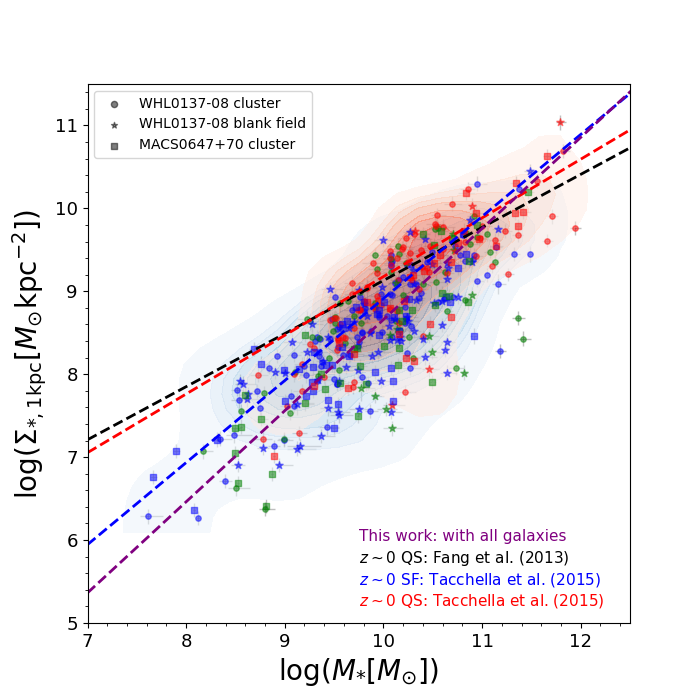}
\caption{A tight relationship between the integrated \mass\ and stellar mass surface density within the central $1$ kpc radius (\sigmakpc). The overall symbols in this figure are the same as those in Figure~\ref{fig:effradSM_vs_effradSFR}. This relation indicates that \sigmakpc\ grows hand-in-hand with the integrated \mass. The quiescent galaxies mostly reside in the top locus of the relation, indicating that they tend to be massive and have a massive spheroid in their centers. The purple dashed line is the best-fit linear function to our sample galaxies, which has a slope of $1.10$, zero-point of $-2.33$, and scatter of $0.38$ dex. The black and red dashed lines are the best-fit to the same relation found in $z\sim 0$ passive galaxies reported by \citet{2013Fang} and \citet{2015Tacchella}, respectively, whereas the blue line is the best-fit to the relation of $z\sim 0$ star-forming galaxies from \citet{2015Tacchella}.}
\label{fig:mass_vs_sigma1kpc}
\end{figure}

Figure~\ref{fig:mass_vs_sigma1kpc} shows a relationship between the integrated \mass\ and \sigmakpc. A tight relationship is evidenced from this figure, which indicates that \sigmakpc\ grows hand-in-hand with the integrated \mass of the galaxies. We fit the relation involving all galaxies in our sample with a linear function (in logarithmic scale) using the Orthogonal Distance Regression (ODR) method and find that the relationship has a scatter of $0.38$ dex. The best-fit linear function (with a slope of $1.10$ and zero-point of $-2.33$) is shown with the purple dashed line. It is interesting to see that quiescent galaxies mostly reside in a tight locus at the top of the relation. This trend indicates that quiescent galaxies tend to be massive and have a massive central stellar mass density, perhaps associated with a bulge that has been formed in these galaxies. This massive central component might in part causes the more compact (i.e.,~smaller half-mass radius) size of quiescent galaxies compared to the star-forming galaxies. In Section~\ref{sec:galaxies_grow_quench}, we will discuss further how this central stellar mass density evolves with redshift and correlates with star formation in the inner and outer regions of the galaxies.  

A similar relationship has also been observed by previous studies in galaxies at $z\lesssim 3$ \citep[e.g.,][]{2013Fang, 2015Tacchella, 2017Barro}. Using a larger sample than ours, they also found that quiescent galaxies occupy a tight locus on top of the overall relationship with all galaxies. The relation involving only quiescent galaxies has a shallower slope than that with star-forming galaxies only. The black and red dashed lines are the best-fit to the \mass--\sigmakpc\ relation of $z\sim 0$ passive galaxies reported by \citet{2013Fang} and \citet{2015Tacchella}, respectively, whereas the blue line is the best-fit to the relation of $z\sim 0$ star-forming galaxies from \citet{2015Tacchella}. In our result, we also see a slight bending at the tip of the distribution of quiescent galaxies, as can be seen from the density contour. It is interesting to see the consistency between the \mass--\sigmakpc\ relation from our study and that from the literature despite the fact that our sample galaxies cover a wider redshift range ($0.3<z<6.0$). It suggests that this relation might be universal and galaxies evolve along this tight relation.

\section{Discussions} \label{sec:discussions}

\subsection{How do Galaxies Grow and Quench Over Cosmic Time?} \label{sec:galaxies_grow_quench}

In this section, we exploit our results to try to infer how galaxies assemble their structures and eventually cease their star formation activities over the course of their life. In particular, we are interested to investigate how the processes of stellar mass buildup and quenching were propagating within the galaxies. Based on the models of the hierarchical galaxy formation paradigm, within the framework of the $\Lambda$CDM cosmology, galaxies are predicted to build their structures over cosmic time in an inside-out manner where the central component was built first and subsequently the disk structure is assembled gradually over time \citep[e.g.,][]{2000Cole,2002vandenBosch,2013Aumer}. This assembly process happens through the series of gas accretions via mergers and filamentary accretion through the cosmic web. 

In a gas-rich merger, gas can fall rapidly into the center of the gravitational potential, causing compaction of gas that eventually triggers a nuclear starburst event \citep[e.g.,][]{2014Dekel}. Other smoother gas streams can also lead to central gas compaction, e.g.,~counter-rotating streams and low angular momentum recycled gas. Based on the zoom-in cosmological hydrodynamical simulations, this wet compaction event is predicted to typically occur in $\sim 10^{9.5}M_{\odot}$ galaxies at $z\sim 2-4$ \citep[e.g.,][]{2015Zolotov,2016Tacchella,2016Tacchella2}. During this event, the star formation at the center is very intense, which converts massive gas concentration rapidly into stars (i.e.,~with short depletion time). This process depletes the gas in galaxies and makes SFR decline. However, the gas compaction event can occur multiple times in the high redshift galaxies, causing an up-and-down of their SFRs. This could appear as an oscillation around the SFMS ridge line. The gas compaction and nuclear starburst processes can build the massive bulge in the centers of the galaxies \citep{2016Tacchella,2016Tacchella2}. Subsequently, galaxies can reach full quenching when the timescale of gas replenishment in the disk is longer than the timescale of gas depletion by star formation. This is more likely to happen at lower redshifts when the cosmological gas accretion rate is low, and especially so in a hot halo above the critical halo mass of $\sim 10^{11.5}M_{\odot}$ \citep{2003Birnboim,2005Keres,2006Dekel}. In general, an emerging picture from the above scenarios is that galaxies build their structures and quench their star formation in an inside-out manner.  

Next, we will exploit our observational results to find indications of the above model predictions. We note that our current sample is limited, which may cause some biases in our interpretations. More information from multiwavelength observations is needed for a more comprehensive study on this, including maps of the gas mass and their kinematics which can be obtained from radio and integral field spectroscopy observations. We leave this for future work.     

\begin{figure*}
\centering
\includegraphics[width=0.65\textwidth]{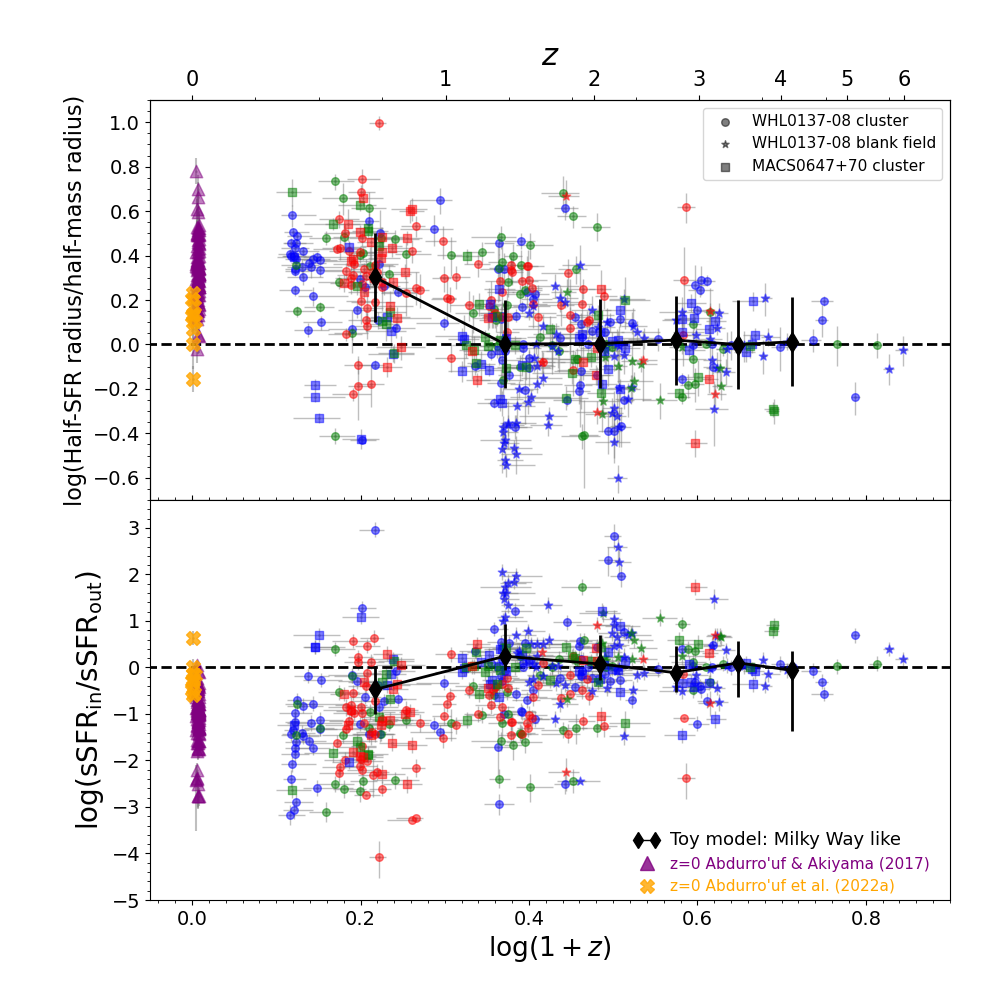}
\caption{Evolution of the ratio between the half-SFR and half-mass radii (top panel) and the ratio between the sSFR inside and outside of the half-mass radius (bottom panel). The overall symbols and color-coding are the same as those in the right panel of Figure~\ref{fig:global_sfms}. The profiles shown with black diamond symbols represent an expectation from a toy empirical model for the evolution of the Milky Way analogs. For comparison, we also show the trends observed in local spiral galaxies from \citet{2017Abdurrouf} and \citet{2022Abdurrouf1}. At $z\gtrsim 3.5$, the two ratios are close to unity, indicating a similar mass doubling time across the galaxy's region. At $1.5\lesssim z\lesssim 2.5$, many star-forming galaxies have a compact star formation (\ratioradius\ as low as $\sim -0.5$ dex and \ratiossfr\ up to $\sim 3$ dex). The majority of the green valley and quiescent galaxies at this epoch have \ratioradius$>1$ and \ratiossfr$<1$. At a later epoch ($z\lesssim 1.5$), the majority of quiescent, green valley, and star-forming galaxies have extended SFR distributions and centrally-suppressed sSFR. Overall these trends point toward the inside-out growth and quenching scenario.}
\label{fig:redshift_vs_ratio_sSFRinout}
\end{figure*}

First, we examine the evolution of the ratio between the half-SFR and half-mass radii and the ratio between the sSFR in the inner (\ssfrin) and outer (\ssfrout) regions of the galaxies. We show the evolution of these quantities in Figure~\ref{fig:redshift_vs_ratio_sSFRinout}. The \ssfrin\ and \ssfrout\ are defined as the total sSFR inside and outside of the half-mass radius, respectively. In the study of galaxy evolution, we face the fact that we observe galaxies at a certain cosmic time. In other words, the galaxies across the wide redshift that we observe here are not necessarily connected evolutionarily (i.e.,~they may not progenitor--descendant pairs), which makes it difficult to interpret an evolutionary trend from our study. Previous studies have tried to connect galaxies using a \mass\ growth function ($M_{*}(z)$) derived from the stellar mass function assuming a constant number density \citep[e.g.,][]{2010vanDokkum,2013vanDokkum}. To try to connect galaxies in our sample and infer some evolutionary trends from their properties, we use $M_{*}(z)$ from \citet{2013vanDokkum} (Eq. 1 therein), which was designed for tracing the evolution of Milky Way analogs (i.e.,~having $M_{*}=5\times 10^{10}M_{\odot}$ at $z=0$). We bin redshift with a width of $0.7$. By using the $M_{*}(z)$ function, We then choose galaxies that fall within $\pm 0.35$ dex in \mass\ at each redshift bin. Our model only extends up to $z=4.5$ because there are no low-mass galaxies selected beyond it in our sample. This can be caused by a bias in our sample selection (see Section~\ref{sec:sample_selection}). This toy empirical model is only intended for a reference in interpreting observational trends. We show the evolutionary trends of this toy model with black diamond symbols. For comparison, we also show the trends observed in local spiral galaxies, as inferred from the data analyzed by \citet{2017Abdurrouf} and \citet{2022Abdurrouf1} (personal communication).  

The ratio between the half-SFR and half-mass radii (\ratioradius) tells about the relative extent of the SFR distribution compared to the stellar mass distribution. A ratio \ratioradius$<1$ implies a more compact SFR distribution than the stellar mass, whereas \ratioradius$>1$ implies a more compact mass distribution than star formation. \ratioradius$<<1$ may indicate an ongoing nuclear starburst, while \ratioradius$>>1$ indicates that a massive bulge has been built in the galaxies. From Figure~\ref{fig:redshift_vs_ratio_sSFRinout}, we can see that from an early epoch up to $z\sim 3.5$, the two ratios are close to unity, indicating a similar mass doubling time across the galaxy's region. In our sample, we only have star-forming and green-valley galaxies at this epoch. Quiescent galaxies emerge from $z\sim 3$ in our sample and we start to see more dispersion in the two ratios at this later epoch. At $1.5\lesssim z\lesssim 2.5$ we see a large fraction of our star-forming galaxies have a compact star formation with \ratioradius\ as low as $\sim -0.5$ dex and centrally-peaked sSFR with \ratiossfr\ of up to $\sim 3$ dex. In contrast to this, the majority of green valley and quiescent galaxies at this epoch have \ratioradius$>1$ and \ratiossfr$<1$. At a later epoch ($z\lesssim 1.5$), the majority of quiescent, green valley, and star-forming galaxies have extended SFR distributions and centrally-suppressed sSFR. The toy model has constant ratios of $\sim 1$ from the early epoch up to $z\sim 1.5$ after which its sSFR declines and SFR distribution becomes more extended. There is an indication that its \ratiossfr\ actually increases a little bit above 1 at $z\sim 1.5$.

The trends observed at $z\sim 0$ indicate that local spiral galaxies have overall extended SFR distributions and centrally-suppressed sSFR radial profiles. These trends agree with the scenarios inferred from our results in the current work and provide a nice extension to our results toward low redshift, complementing the general picture.    

Overall, the above trend agrees with the inside-out growth and quenching scenarios. In the early cosmic time, galaxies get steady gas accretion for star formation but they have yet to form a bulge, and the star formation is likely distributed evenly across their regions. At $z\sim 2$, which coincides with the peak epoch of the cosmic SFRD and perhaps the cosmic gas accretion \citep{2014Madau}, star-forming galaxies in our sample may experience gas compaction events that later build bulge in their centers. After that, quenching might have been started in their central regions, but star formation is still active in the disk that further building the disk. In addition to in-situ star formation, minor mergers can also contribute to the buildup of stellar mass in the disk and grow the galaxy size.

\subsection{The buildup of the Central Stellar Mass Density Over Cosmic Time} \label{sec:evolve_centmass}

\begin{figure*}
\centering
\includegraphics[width=0.65\textwidth]{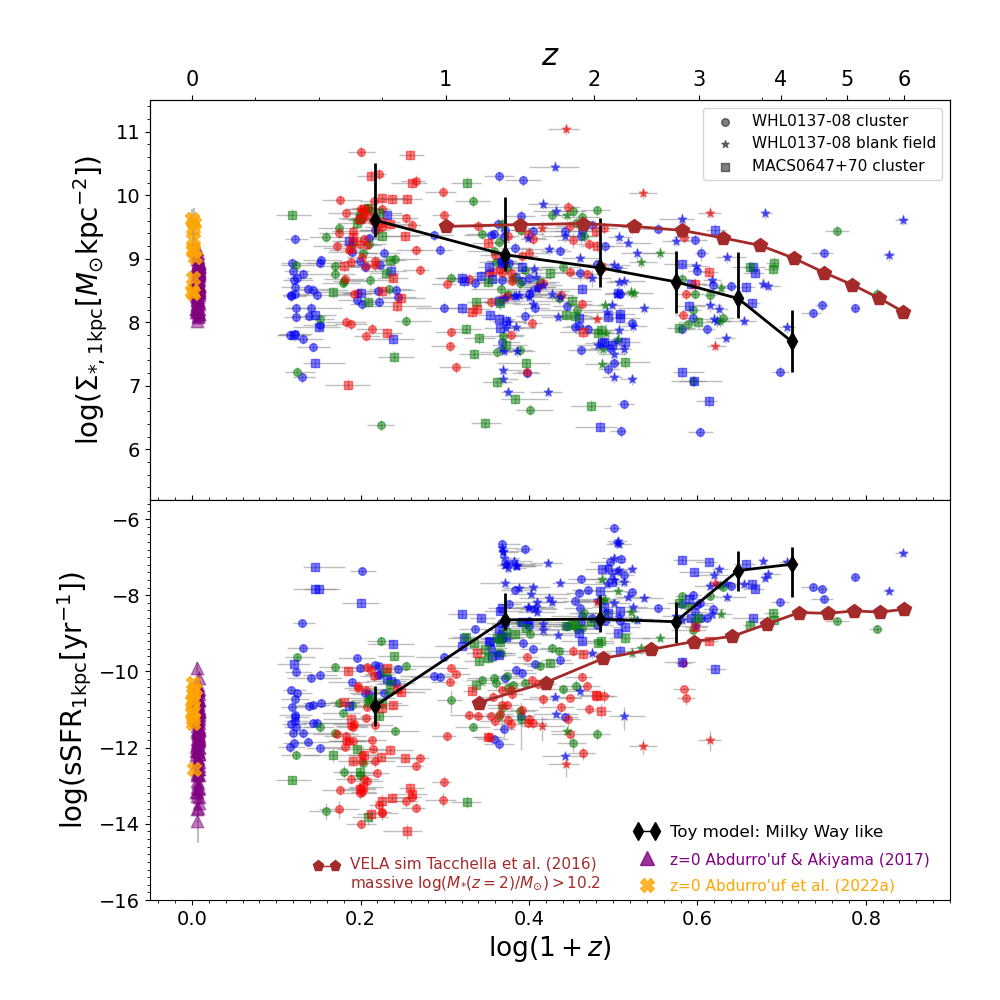}
\caption{The buildup of the stellar mass density at the central 1 kpc (\sigmakpc) over cosmic time (top panel) and the evolution of sSFR at the central 1 kpc (\ssfrkpc; bottom panel). The profiles shown with black diamond symbols represent an expectation from a toy empirical model for the evolution of the Milky Way analogs, whereas the red pentagon symbols show a prediction from zoom-in cosmological hydrodynamical simulations by \citet{2016Tacchella2} for the evolution of massive galaxies ($\log(M_{*}(z=2)/M_{\odot})>10.2$).}
\label{fig:redshift_vs_sigma1kpc}
\end{figure*}

As we have seen in Section~\ref{sec:cent_mass_sd}, our sample galaxies exhibit a tight relationship between the global \mass\ and \sigmakpc. \sigmakpc\ is a good indicator for quiescent galaxies because they form a rather distinct sequence at the tip of the overall $M_{*}$--\sigmakpc\ relation and have a shallower slope. Here we discuss the evolution of \sigmakpc\ with redshift to see how the central bulge is built over cosmic time in our sample galaxies. As \sigmakpc\ develops over time, it is also interesting to analyze how sSFR at the central 1 kpc evolves following the development of \sigmakpc. The evolution of these two quantities is shown in Figure~\ref{fig:redshift_vs_sigma1kpc}. As we can see from this figure, \sigmakpc\ tends to increase with cosmic time, whereas \ssfrkpc\ declines with cosmic time. The quiescent galaxies tend to have higher \sigmakpc\ and lower \ssfrkpc\ in all redshifts. \ssfrkpc\ of quiescent and green-valley galaxies tend to be declined more rapidly than that of star-forming galaxies. Interestingly, the overall \ssfrkpc\ of our star-forming galaxies does not decline much from $z=6$ up to $z\sim 1.5$. There is an indication that \ssfrkpc\ of some star-forming galaxies even increases at $1.5\lesssim z\lesssim 2.5$.                 

The black profiles show the expected evolution of Milky Way analogs based on our toy model derived in Section~\ref{sec:galaxies_grow_quench}. Its \sigmakpc\ increases by $\sim 2$ magnitude over $0.5\lesssim z\lesssim 4.5$, while its \ssfrkpc\ decreases significantly ($\sim 3.7$ magnitude) over the same period. This implies that the central SFR within 1 kpc also decreases with time. At $1.5\lesssim z\lesssim 2.5$, the \ssfrkpc\ of this model seems to be constant. In addition to our toy model, we also compare our observational trend with the predictions from the VELA zoom-in cosmological hydrodynamical simulations \citep{2014Ceverino,2015Zolotov} that was analyzed by \citet{2016Tacchella2}. These predictions, which are shown in red profiles, are obtained by averaging the evolutionary trends of relatively massive galaxies in the simulations that have $\log(M_{*}/M_{\odot})=10.2$ at $z=2$. This simulation was run over $1<z<7$. 
The \sigmakpc\ and \ssfrkpc\ predicted from the cosmological simulation at $z\sim 6$ are in good agreement with our observations. The evolution of the \sigmakpc\ and \ssfrkpc\ from the simulation might be consistent with the evolution of the progenitors of local massive quiescent galaxies, as implied from our observations. However, there is an excess of \sigmakpc\ at $z\sim 0.5$ quiescent galaxies compared to the simulation. Those galaxies are likely members of the \WHL\ or \MACS\ clusters. In clusters, galaxies could accrete more mass through mergers which can result in denser central mass density. 

The trends at $z\sim 0$ from \citet{2017Abdurrouf} and \citet{2022Abdurrouf1} provide a good extension for our results in the current work. They overall agree with the picture of increasing \sigmakpc\ and decreasing \ssfrkpc\ with cosmic time. However, we also see a lower \sigmakpc\ in the local spiral galaxies than in the quiescent galaxies at $z\sim 0.5$ that are possibly the cluster members. If we ignore these galaxies and assume that \sigmakpc\ trend from the VELA simulation will evolve to have a similar value as those of the observed \sigmakpc\ of local galaxies (which is likely given its shallow slope at $z\sim 1$), we see a possible saturation of central mass density in galaxies.         

\begin{figure*}
\centering
\includegraphics[width=0.8\textwidth]{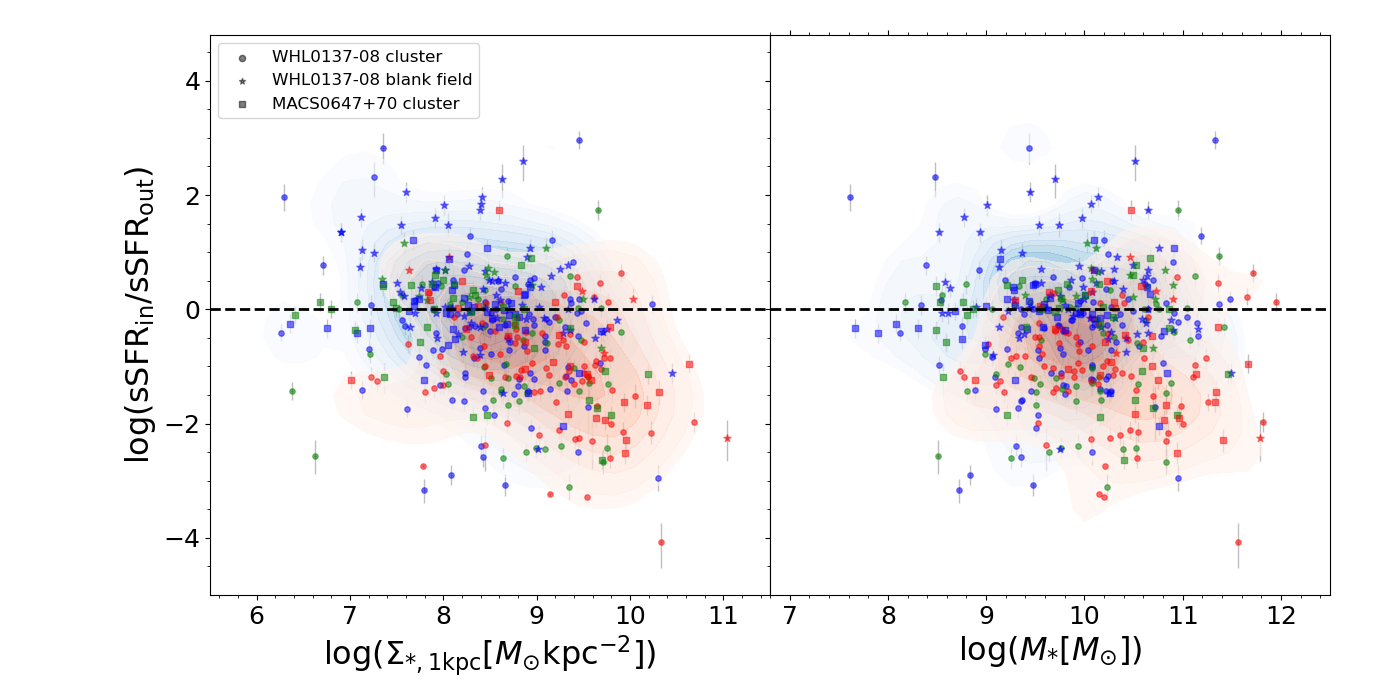}
\caption{The \ratiossfr\ ratio as a function of the central mass density (left panel) and global \mass (right panel). The overall symbols in this figure are the same as those in Figure~\ref{fig:effradSM_vs_effradSFR}. Galaxies that are massive and have high \sigmakpc\ tend to have centrally-suppressed sSFR profiles. This trend is predominantly observed in quiescent galaxies. The \mass--\ratiossfr\ relation seems to be broader and less significant compared to \mass--\ratiossfr, indicating that \sigmakpc\ is more influential in driving \ratiossfr\ than global \mass.}
\label{fig:sigma1kpc_vs_sSFRinout}
\end{figure*}

Next, we compare the effects of \sigmakpc\ and the global $M_{*}$ on \ratiossfr. We show the \sigmakpc--\ratiossfr\ and \mass--\ratiossfr\ relations in Figure~\ref{fig:sigma1kpc_vs_sSFRinout}. We can see from this figure that \ratiossfr\ is correlated with both \sigmakpc\ and the global $M_{*}$ such that increasing \sigmakpc\ and $M_{*}$ corresponds to a steeper sSFR decline in the central regions. However, the \mass--\ratiossfr\ relation seems to be broader and less significant compared to \mass--\ratiossfr, indicating that \sigmakpc\ is more influential in driving \ratiossfr\ than global \mass. The majority of those that have central sSFR suppression are the quiescent and green-valley galaxies, whereas a significant fraction of star-forming galaxies have broadly flat or centrally-peaked sSFR radial profile (i.e.,~negative gradient). This further suggests that \sigmakpc\ is a good predictor for quiescent galaxies, which agree with previous studies \citep[e.g.,][]{2013Fang, 2012Cheung, 2017Jung, 2017Barro, 2017Whitaker, 2022Bluck}.           

Some physical mechanisms for the quenching in galaxies have been proposed, including the stellar feedback \citep[e.g.,][]{1986Dekel, 2005Murray}, AGN feedback \citep[e.g.,][]{2005DiMatteo, 2006Croton, 2007Ciotti, 2009Cattaneo}, kinematic stabilization of gas in the disk by massive bulge \citep[e.g.,][]{2009Martig, 2014Genzel}, and a long-term suppression of gas supply due to the virial shock heating by the dark matter halo when it has reached a threshold mass of $\sim 10^{11.5}M_{\odot}$ \citep[e.g.,][]{2003Birnboim, 2006Dekel, 2009Keres}. The inside-out quenching observed in this work seem to agree with the AGN feedback scenario, as has also been suggested by previous spatially resolved studies of galaxies at $z\lesssim 2$ \citep[e.g.,][]{2022Bluck, 2021Nelson}. The feedback from AGN can expel or heat up gas, which then cause the suppression of star formation in at least the central region of the galaxies. The feedback strength must scale with the mass of the supermassive black hole (SMBH), which has been known to correlate tightly with the bulge mass \citep[e.g.,][]{2004Haring, 2019Schutte}. This is overall in line with our results which show that quiescent galaxies tend to have centrally suppressed sSFR and high central mass density. However, more in-depth study is needed to further investigate this. Ideally, we need multi-wavelength data sets that allow us to identify AGN and measure the strength of its feedback, which we have not had in this study.

\section{Summary and Conclusions} \label{sec:conclusions}

We perform spatially resolved SED fitting on 444 galaxies at $0.3<z<6.0$ in two clusters (\WHL\ and \MACS) and a blank field using imaging data from \JWST\ and \HST\ in up to 13 bands. We use \pixedfit\ throughout the analysis. This software can simultaneously perform image processing, pixel binning, and spatially resolved SED fitting. By using the maps of spatially resolved stellar population properties (on kpc scales) obtained from this analysis, we investigate how galaxies grow their structures and quench their star formation activities across cosmic time. Overall, our key results are summarized in the following:
\begin{enumerate}
\item  The normalization of the stellar mass surface density radial profiles (\sigmakpc$(r)$) increases with increasing cosmic time and global \mass. At each redshift, quiescent galaxies tend to have higher \sigmakpc\ across the entire radius than green-valley and star-forming galaxies. The sSFR radial profiles (sSFR$(r)$) show more variations across redshift and global \mass. The sSFR$(r)$ are broadly flat at $2.5\lesssim z\lesssim 6.0$ in all galaxies, indicating a similar mass-doubling time across the entire radius. At $0.8\lesssim z\lesssim 2.5$, less massive ($\log(M_{*}/M_{\odot})<11.0$) star-forming galaxies have flat or centrally-peaked sSFR$(r)$, whereas the majority of quiescent galaxies have centrally-suppressed sSFR$(r)$. At lower redshift ($z<0.8$), almost all galaxies (regardless of \mass\ and star formation stage) have centrally-suppressed sSFR$(r)$. The radial profiles of stellar ages show that those galaxies with centrally-peaked sSFR$(r)$ at $0.8\lesssim z\lesssim 2.5$ have very young stellar populations in their central regions, indicating an ongoing nuclear starburst.

\item The majority of quiescent galaxies have a larger half-SFR radius than the half-mass radius, indicating that they have extended spatial distribution of SFR and compact distribution of stellar mass. In contrast, some star-forming galaxies, especially at high redshifts, have a half-SFR radius being roughly similar or smaller to the half-mass radius, whereas those at low redshifts have a half-SFR radius being larger than the half-mass radius. The half-mass radius of the star-forming galaxies is on average larger than the quiescent galaxies in all global \mass.  

\item We observe a tight correlation between the global \mass\ and stellar mass density at the central 1 kpc radius (\sigmakpc) with $0.38$ dex, indicating that galaxies grow their central mass density hand-in-hand with their global \mass. The quiescent galaxies reside in a sequence at the tip of the overall relationship and have a shallower slope. This trend indicates that \sigmakpc\ is a good predictor of quenching, where passive galaxies tend to have higher \sigmakpc\ and global \mass. The shallower slope of \mass--\sigmakpc\ in quiescent galaxies suggests that their central mass density has reached a saturation point.

\item We investigate the evolution of the \ratioradius\ and \ratiossfr\ ratios with redshift to try to understand how galaxies grow their structures and quench their star formations over cosmic time. We find that the ratios are close to unity from the early epoch up to $z\sim 3.5$ and the ratios start to deviate from unity since then. At $1.5\lesssim z\lesssim 2.5$, a fraction of our star-forming sample has a low \ratioradius\ and high \ratiossfr, indicating that they may be experiencing a nuclear starburst. At the later epoch, most of our sample galaxies, especially quiescent and green-valley have a high \ratioradius\ and low \ratiossfr, suggesting that massive bulges might have been formed in these galaxies and the star formation has been quenched in their central regions.

\item We also investigate the evolution of \sigmakpc\ and sSFR at the central 1 kpc (\ssfrkpc). In general, we see an increasing \sigmakpc\ and decreasing \ssfrkpc\ with cosmic time, indicating the buildup of the central bulge component and the quenching process in the central region of the galaxies. We also find that quiescent galaxies tend to have higher \sigmakpc\ and lower \ssfrkpc\ than star-forming galaxies in all redshifts.

\item Finally, we observe \sigmakpc--\ratiossfr\ and \mass--\ratiossfr\ relations with a negative slope, indicating that galaxies that are more massive and have massive \sigmakpc\ tend to have steeper sSFR suppression in their centers. The \sigmakpc--\ratiossfr\ relation seem to be tighter than \mass--\ratiossfr\, indicating that \sigmakpc\ is more influential in driving \ratiossfr\ than global \mass. The quiescent galaxies tend to have higher \sigmakpc\ and \ratiossfr$<1$, suggesting that the formation of bulge might happen simultaneously with the quenching of star formation in the central regions.     
\end{enumerate}

Our work in this paper demonstrates the great potential of spatially resolved SED analysis using \JWST\ imaging data. It is interesting to extend this study with larger sample galaxies taken from various surveys to better understand the buildup of stellar mass in galaxies and the growth of their structures over cosmic time. More comprehensive comparisons with zoom-in cosmological simulations would help to better understand the underlying physics. We will pursue this in our future work.  



\section*{Acknowledgements}

We thank the anonymous referee for providing valuable comments that help to improve this paper. We thank Takahiro Morishita for the useful discussion and comments. This work is based on observations made with the NASA/ESA/CSA James Webb Space Telescope (JWST). Some of the data presented in this paper were obtained from the Mikulski Archive for Space Telescopes (MAST) at the Space Telescope Science Institute. The specific observations analyzed can be accessed via \dataset[http://dx.doi.org/10.17909/d2er-wq71]{http://dx.doi.org/10.17909/d2er-wq71} and \dataset[http://dx.doi.org/10.17909/cqfq-5n80]{http://dx.doi.org/10.17909/cqfq-5n80}. STScI is operated by the Association of Universities for Research in Astronomy, Inc., under NASA contract NAS5–26555. Support to MAST for these data is provided by the NASA Office of Space Science via grant NAG5–7584 and by other grants and contracts.

A and TH are funded by a grant for JWST-GO-01433 provided by STScI under NASA contract NAS5-03127. The CosmicDawn Center is funded by the Danish National Research Foundation (DNRF) under grant \#140. PD acknowledges support from the NWO grant 016.VIDI.189.162 (``ODIN") and from the European Commission's and University of Groningen's CO-FUND Rosalind Franklin program. 
RAW acknowledges support from NASA JWST Interdisciplinary Scientist grants NAG5-12460, NNX14AN10G and 80NSSC18K0200 from GSFC. AZ and AKM acknowledge support by grant 2020750 from the United States-Israel Binational Science Foundation (BSF) and grant 2109066 from the United States National Science Foundation (NSF), and by the Ministry of Science \& Technology, Israel. MO acknowledges support from JSPS KAKENHI Grant Numbers JP22H01260, JP20H05856, JP20H00181, JP22K21349. AA acknowledges support from the Swedish Research Council (Vetenskapsr\aa{}det project grants 2021-05559). EV acknowledges financial support through grants PRIN-MIUR 2017WSCC32, 2020SKSTHZ and the INAF GO Grant 2022 (P.I. E. Vanzella).

%

\vspace{5mm}
\facilities{\textit{HST}, \textit{JWST}}


\software{\astropy\ \citep{2013Astropy,2018Astropy,2022Astropy}, 
\pixedfit\ \citep{2021Abdurrouf, 2022Abdurrouf3}, \sextractor\ \citep{1996Bertin}, \SEP\ \citep{SEP}, \photutils\ \citep{Bradley2022b}, \grizli\ \citep{Grizli}, \eazypy\ \citep{2008Brammer}. 
}



\appendix

\section{Robustness of the SED Fitting Method: Fitting Tests with Mock SEDs} \label{sec:fit_tests}

To test the robustness of our SED fitting method on this new set of photometric data, we perform SED-fitting tests using semiempirical mock SEDs, following a similar procedure as performed in \citet[][Appendix A therein]{2022Abdurrouf1}. We draw the parameter values for our mock SEDs from the measured parameters of real galaxies. In this case, we use the measured parameters obtained from our fitting to the SEDs within the central effective radius that were used for determining our photometric redshifts (see Section~\ref{sec:resolved_sedfit}). Here, we use 290 galaxies selected randomly from 354 galaxies in the \WHL\ (before further exclusion). We prefer to use the parameters of real galaxies for generating mock SEDs, instead of drawing them randomly because we can not be sure that the combinations of those random parameters are physically realistic. Using the set of parameters of 290 galaxies, we generate mock SEDs using the same modeling setup as in our main analysis. We generate two sets of mock SEDs. The first one with 12 filters of \JWST\ and \HST, the same set of filters as available for the \WHL\ cluster. The second set with 8 \JWST\ filters, excluding \HST\ filters. We then inject Gaussian noises assuming S/N of 20 in all filters. After that, we fit the mock SEDs using the same method as we used in the main analysis of this paper. For simplicity, here we fix redshift. 

\begin{figure*}
\includegraphics[width=0.33\textwidth]{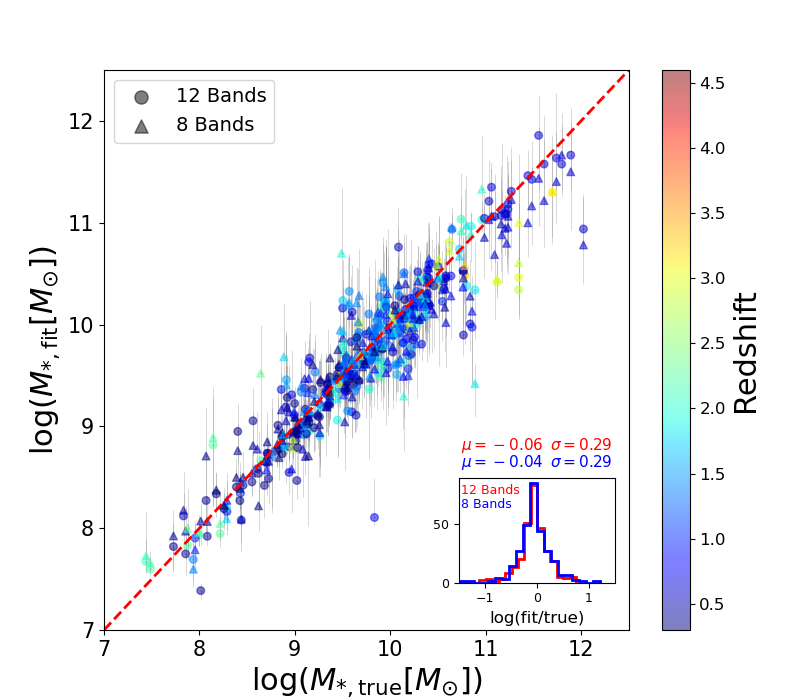}
\includegraphics[width=0.33\textwidth]{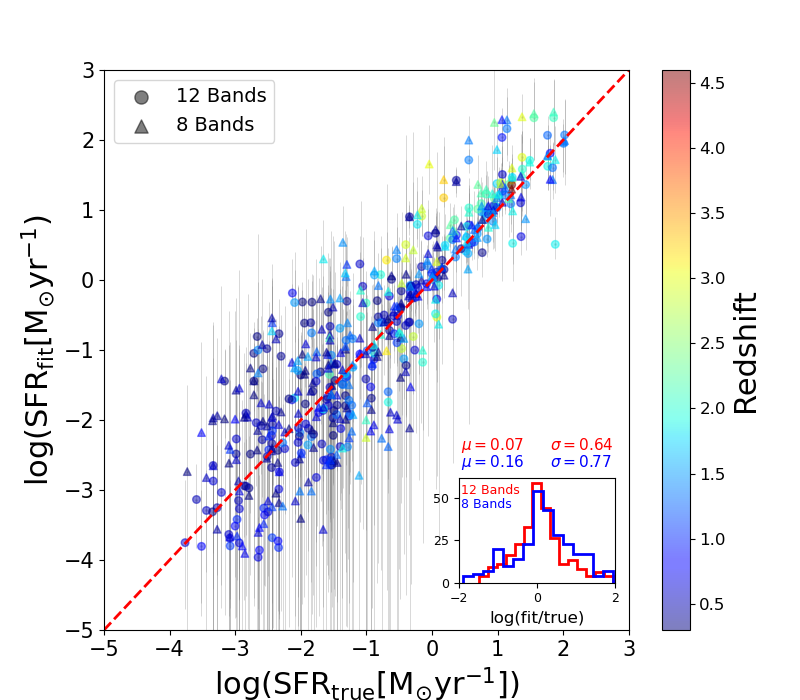}
\includegraphics[width=0.33\textwidth]{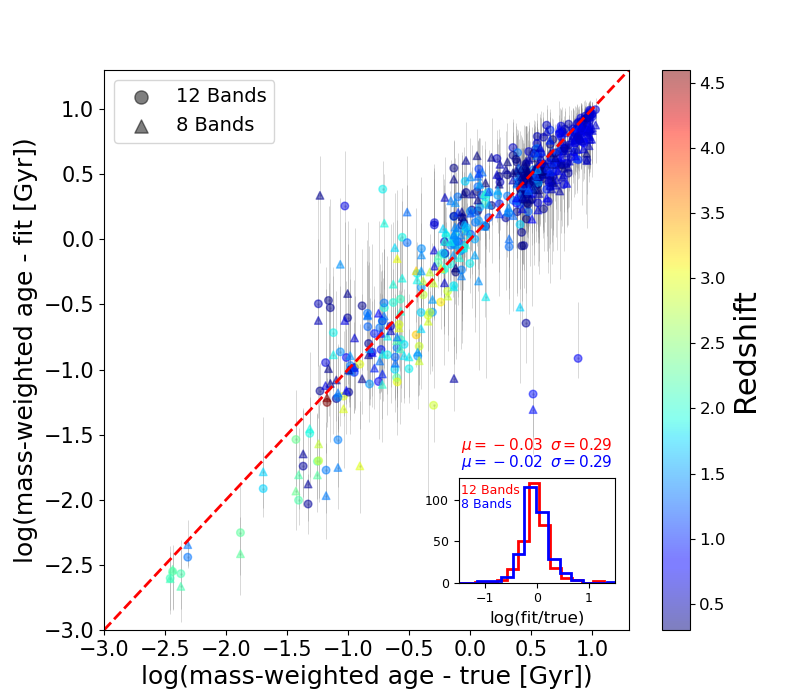}
\includegraphics[width=0.33\textwidth]{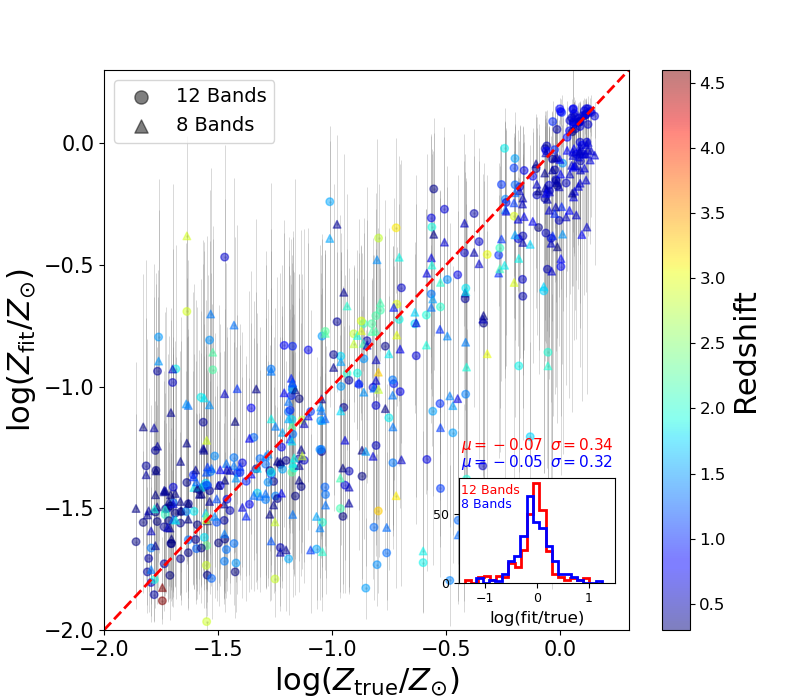}
\includegraphics[width=0.33\textwidth]{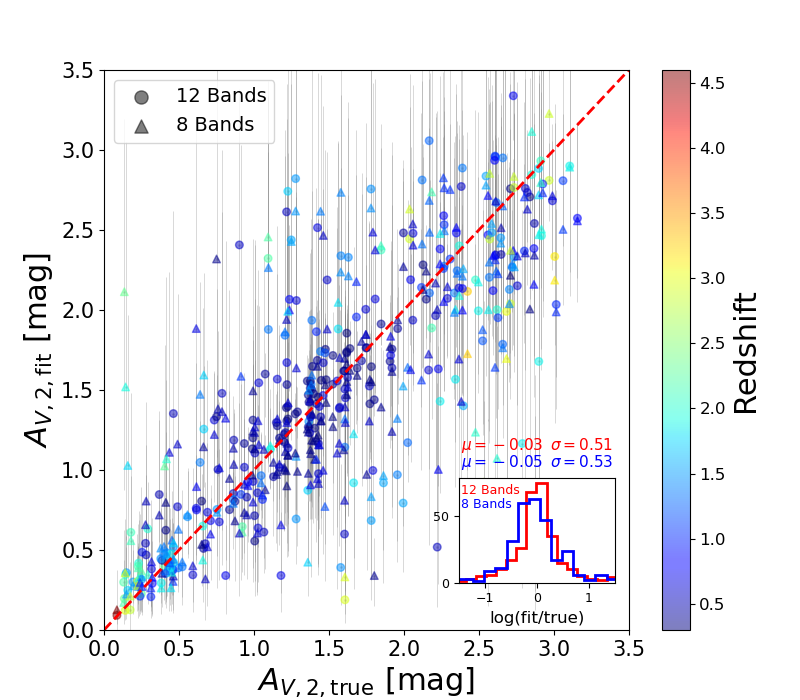}
\caption{Comparisons between the best-fit parameters derived from SED-fitting tests using mock SEDs and the ground truth. In this test, two sets of photometry are used, one with combined 12 filters of \HST\ and \JWST\ and the other with 8 filters of \JWST\. The results of those two SED fitting tests are shown with different symbols (circles and triangles). Histograms in the insets show ratios between the best-fit parameters and the true values from the mock SEDs. The color-coding of the data points represents redshift.}
\label{fig:fitting_tests}
\end{figure*}

We present the results in Figure~\ref{fig:fitting_tests}, which shows the comparisons between the best-fit parameters derived from SED fitting and the true values from the mock SEDs. Histograms in the insets show ratios between the best-fit parameters and the true values. Results of the SED fitting with two sets of photometry are shown with different symbols and the histograms are shown with different colors. The data points are color-coded based on their redshifts. 

Overall, our SED fitting can recover the true parameters reasonably well. Stellar mass and mass-weighted age are recovered very well in both 8-band and 12-band SED fitting with small offset ($\lesssim 0.06$ dex) and small standard deviation ($\lesssim 0.3$ dex). It is interesting to see that mass-weighed age is well recovered here. This may be due to the fact that our photometry covers the Balmer break and we have good photometry in rest-frame NIR from \JWST, which also provides a good constraint for \mass\ (see Appendix~\ref{sec:breaking_degeneracies}). The stellar metallicity ($Z$) and dust attenuation in the diffuse ISM ($A_{V,2}$) are also recovered well with an offset of $\lesssim 0.07$ dex and a standard deviation of $\lesssim 0.5$ dex although they look to be more scattered due to its small dynamical range. The SFR is more difficult to be recovered for passive galaxies ($\log(\text{SFR}) \lesssim -1$) than for star-forming ones. For the whole sample, the SED fitting with 12 bands gives better SFR estimates (a small offset of $0.07$ dex and scatter of $0.64$ dex) than with only \JWST\ bands (offset of $0.16$ dex and scatter of $0.77$ dex).               

\section{The Age--Dust--Metallicity Degeneracy} \label{sec:breaking_degeneracies}

The addition of \JWST\ NIRCam data extends the wavelength coverage up to roughly the rest-frame NIR for our sample galaxies. This sufficiently wide wavelength coverage has the potential to break the well-known age--dust--metallicity degeneracy in SED fitting, which is very important for the analysis in this paper. In particular, it is crucial to be able to determine if the reddening observed in the central region of some galaxies in our sample is due to aging (i.e.,~quiescence) or dust attenuation. To check if our data set provides a sufficient constraint for resolving this degeneracy, we examine model color-color diagrams among the NIRCam filters. We base our analysis on the observed frame, instead of the rest-frame, and explore different sets of filters for different redshifts. To generate model SEDs, we use an overall similar setting as that used in the main analysis of this paper and assume a double power-law SFH with $\alpha=3.0$, $\beta=0.5$, and $\tau=1.5$ Gyr, and $M_{*}=10^{10.5}M_{\odot}$. We then generate model SEDs in grids of age, $A_{V,2}$, and metallicity. For simplicity, we assume $A_{V,1}=1.5\times A_{V,2}$. 

\begin{figure*}
\centering
\includegraphics[width=1.0\textwidth]{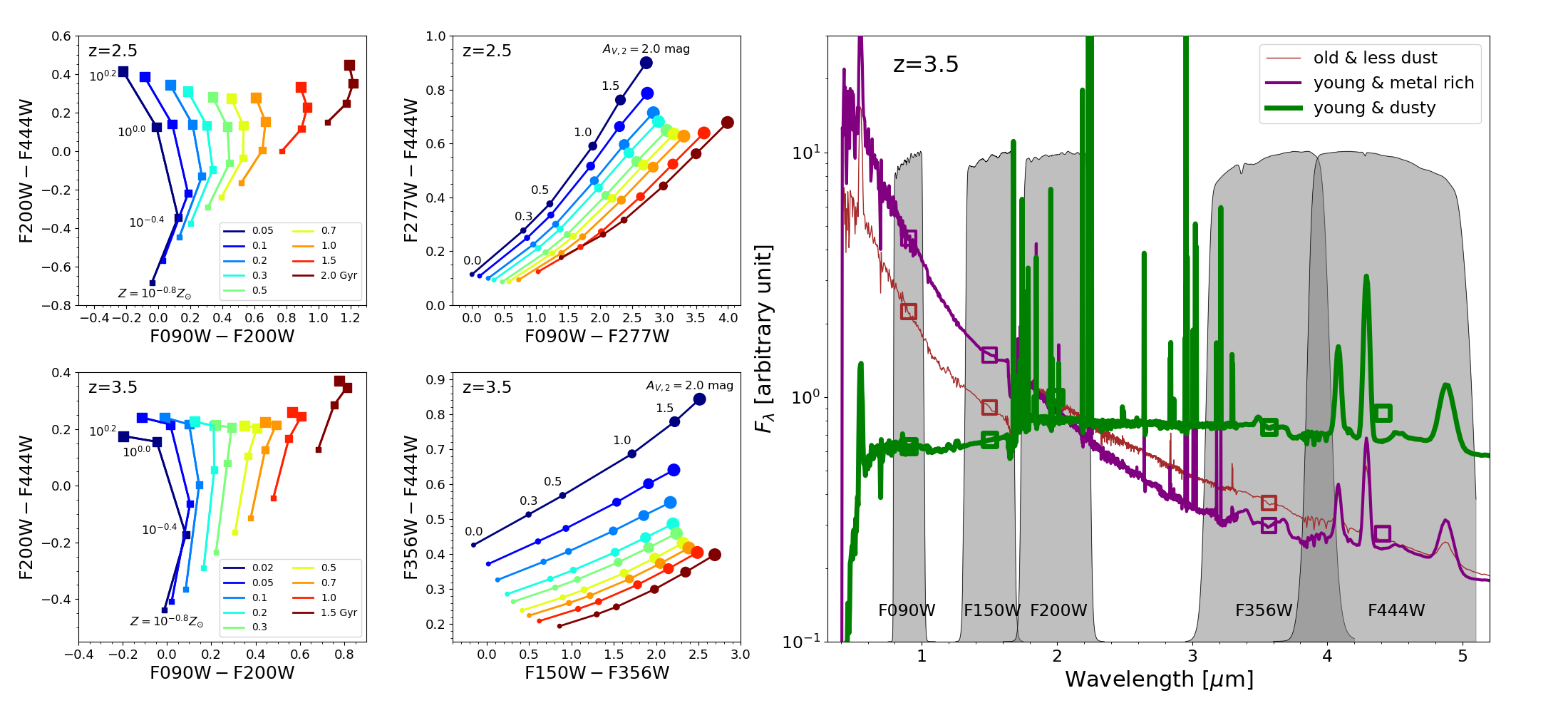}
\caption{The patterns of model stellar populations on the observed-frame color-color diagrams involving \JWST\ NIRCam filters. We plot the patterns of models at $z=2.5$ and $z=3.5$ on the color-color diagrams to check how good the NIRCam photometry is in resolving the degeneracies among age, dust attenuation, and metallicity in high redshifts. Different colors represent different ages, whereas an increasing marker size represents an increasing $Z$ (for the square symbol) and increasing $A_{V,2}$ (for the circle symbol). The rightmost panel shows examples of the model SEDs of old less-dusty (brown color), young metal-rich (purple), and young dusty (green) galaxies at $z=3.5$. The reddening effects by aging and dust are almost orthogonal with each other (i.e.,~distinguishable) on the F090W$-$F200W vs. F200W$-$F444W diagram. The reddening effects by aging and dust attenuation are distinguishable on the F090W$-$F277W vs. F277W$-$F444W for $z=2.5$ and the F150W$-$F356W vs. F356W$-$F444W for $z=3.5$.}
\label{fig:color2_diagram}
\end{figure*}

Figure~\ref{fig:color2_diagram} shows the color-color diagrams of models at $z=2.5$ and $3.5$. Different colors represent different ages, whereas increasing marker size represents increasing metallicity (for the square symbol in the first column) and $A_{V,2}$ (for the circle symbol in the second column). For the two plots in the first column, we fix $A_{V,2}=0.0$, while for the two plots in the second column, we fix $Z=Z_{\odot}$. F090W$-$F200W vs. F200W$-$F444W diagram at the two redshifts seems to be able to distinguish the reddening effect by age and metallicity in such a way that both effects are almost orthogonal with each other. An increasing $Z$ at a fixed age corresponds to reddening in F200W$-$F444W and roughly constant F090W$-$F200W (i.e.,~vertical shift on the diagram). On the other hand, an increasing age tends to make reddening in the two colors (i.e.,~a diagonal shift on the diagram) for galaxies with low $Z$, while it corresponds to a roughly horizontal shift for galaxies with high $Z$. 

In the second column, we show a relative effect of the dust attenuation and aging on F090W$-$F277W vs. F277W$-$F444W (for $z=2.5$) and F150W$-$F356W vs. F356W$-$F444W (for $z=3.5$) diagrams. The two effects are distinguishable in these two diagrams with dust attenuation seems to make a diagonal shift, whereas an aging effect is roughly orthogonal to it. For an illustration, in the rightmost panel, we show examples of the model SEDs of an old less-dusty (brown color; age$1.5$ Gyr, $\log(Z/Z_{\odot})=-0.5$, $A_{V,2}=0.1$ mag), young metal-rich (purple; age$0.1$ Gyr, $\log(Z/Z_{\odot})=0.2$, $A_{V,2}=0.1$ mag), and young dusty (green; age$0.1$ Gyr, $\log(Z/Z_{\odot})=-0.5$, $A_{V,2}=2.0$ mag) galaxies $z=3.5$. We normalize the SEDs by dividing them with the F200W flux. We also show the transmission curves of the filters that are used in the color-color diagrams for $z=3.5$ in the left two columns. As we can see from this figure, the three model SEDs are distinguishable with NIRCam photometry. The reddening due to the dust attenuation is easily recognizable in the rest-frame UV to NIR, while due to the metallicity is not easily recognizable in the rest-frame UV colors but it is detectable in the rest-frame around the Balmer break ($\sim 4000$~\AA) and NIR. 

\section{Construction of empirical Point Spread Functions and Convolution Kernels} \label{sec:generate_psfs}

\begin{figure*}[ht]
\centering
\includegraphics[width=0.4\textwidth]{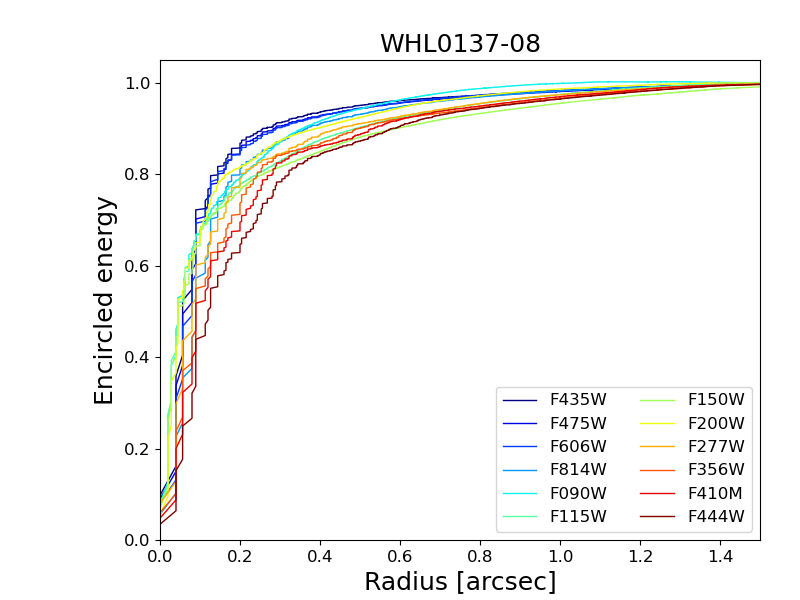}
\includegraphics[width=0.4\textwidth]{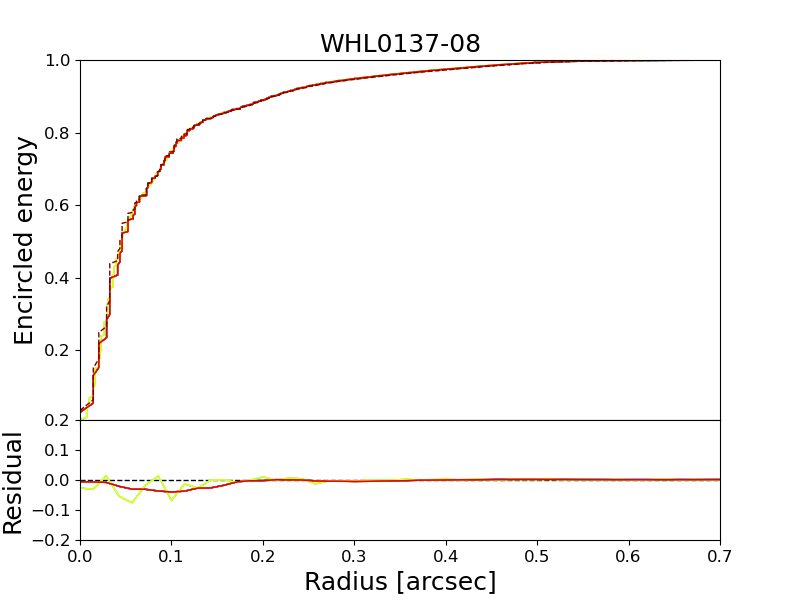}
\includegraphics[width=0.4\textwidth]{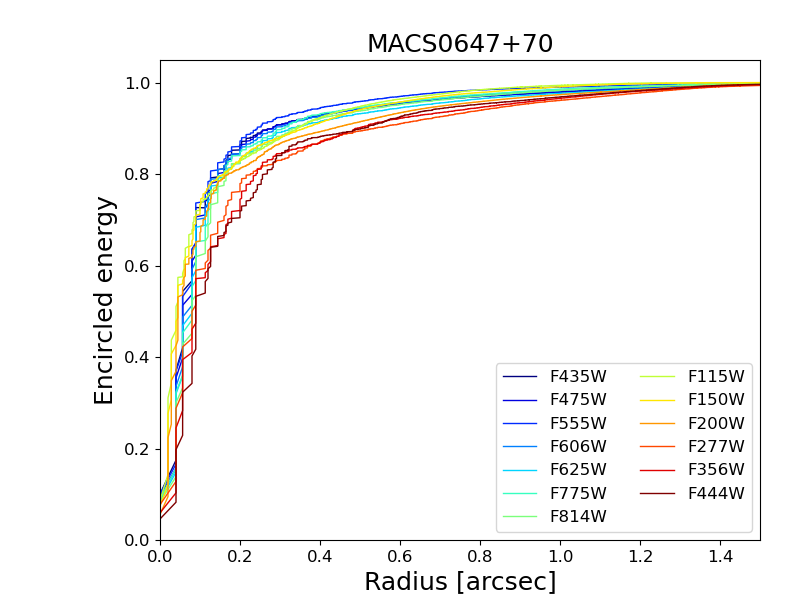}
\includegraphics[width=0.4\textwidth]{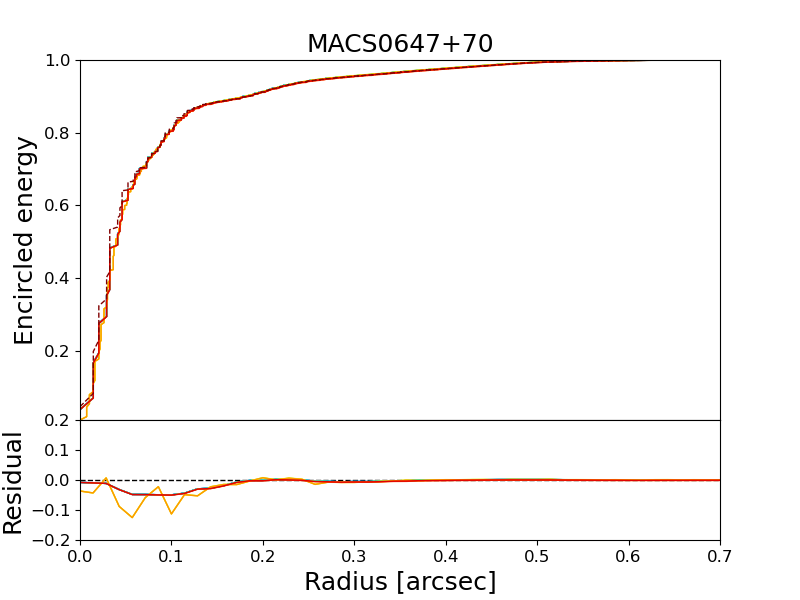}
\caption{\textit{Left column}: Encircled energy of the empirical PSFs of the \HST\ ACS and \JWST\ NIRCam filters in \WHL\ and \MACS\ clusters. The PSFs are generated by stacking the images of bright isolated stars in the fields. \textit{Right column}: Encircled energy of the PSFs after being convolved with kernels. The convolved PSFs have similar encircled energy, indicating the robustness of our PSF matching process.}
\label{fig:emp_PSFs_energy}
\end{figure*}

We generate the empirical PSFs of the \HST\ ACS and \JWST\ NIRcam filters in the \WHL\ and \MACS\ clusters by stacking images of bright isolated stars in those fields. For this, we use \photutils\ package \citep{Bradley2022b}. We show the encircled energy of the empirical PSFs in the left column of Figure~\ref{fig:emp_PSFs_energy}. After generating the PSFs, we then construct the convolution kernels that can be used for PSF matching. As described in Section~\ref{sec:images_analysis}, we perform PSF matching to homogenize the PSF sizes of our imaging data to match the PSF size of F444W, which is the largest among the filters used in our work. We also use \photutils\ for generating the kernels. To check the reliability of our kernels and PSF matching, we convolve the PSF images of the filters other than F444W with the kernels and compare the encircled energy of the convolved-PSFs to that of F444W filter. We show this comparison in the right column of Figure~\ref{fig:emp_PSFs_energy}. As we can see from this figure, the convolved PSFs have similar encircled energy (with a small deviation $<0.1$ dex around a radius of $0\farcs1$), indicating the robustness of our PSF matching process.

\bibliography{resolvedSEDs}{}
\bibliographystyle{aasjournal}



\end{document}